\documentclass[MSNbibl,nameyear,dvips]{arxstspdf}
\usepackage{dcolumn}
\usepackage{graphicx}
\usepackage{url,breakurl}
\usepackage{flushend}
\usepackage{stfloats}

\volume{29}
\issue{4}
\pubyear{2014}
\firstpage{640}
\lastpage{661}
\doi{10.1214/13-STS450} 

\makeatletter
\newcolumntype{d}[1]{D{.}{.}{#1}}
\newcommand{\rrvert}{\vert}
\def\independent{{\perp\!\!\!\!\perp}}
\renewcommand{\citep}[1]{\citeauthor{#1}, \citeyear{#1}}
\newcommand{\iQ}{\textit{Q}}
\newcommand{\iA}{\textit{A}}
\newcommand{\mA}{\mathcal{A}}
\newcommand{\mY}{\mathcal{Y}}
\newcommand{\mD}{\mathcal{D}}
\newcommand{\barS}{\bar{S}}
\newcommand{\barA}{\bar{A}}
\newcommand{\bard}{\bar{d}}
\newcommand{\baru}{\bar{u}}
\newcommand{\barmA}{\bar{\mathcal{A}}}
\newcommand{\barmS}{\bar{\mathcal{S}}}
\newcommand{\mH}{\mathcal{H}}
\newcommand{\mK}{\mathcal{K}}
\newcommand{\mS}{\mathcal{S}}
\newcommand{\tilV}{\widetilde{V}}
\newcommand{\pr}{\mathrm{pr}}

\newcommand{\hatxi}{\widehat{\xi}}
\newcommand{\hatpsi}{\widehat{\psi}}

\newcommand{\logit}{\operatorname{logit}}
\newcommand{\expit}{\operatorname{expit}}
\newcommand{\E}{\mathrm{E}}

\makeatother

\begin{document}
\begin{frontmatter}

\title{$Q$- and $A$-Learning~Methods~for Estimating Optimal Dynamic Treatment~Regimes}
\runtitle{\textit{Q}- and \textit{A}-learning Methods for Optimal Regimes}

\begin{aug}
\author[A]{\fnms{Phillip J.} \snm{Schulte}\ead[label=e1]{phillip.schulte@duke.edu}},
\author[B]{\fnms{Anastasios A.} \snm{Tsiatis}\ead[label=e2]{tsiatis@ncsu.edu}},
\author[C]{\fnms{Eric B.} \snm{Laber}\ead[label=e3]{eblaber@ncsu.edu}}
\and
\author[D]{\fnms{Marie} \snm{Davidian}\corref{}\ead[label=e4]{davidian@ncsu.edu}}
\runauthor{Schulte, Tsiatis, Laber and Davidian}
\affiliation{Duke Clinical Research Institute, North Carolina State University,
North Carolina State University and North Carolina State University}
\address[A]{Phillip J. Schulte is Biostatistician, Duke Clinical
Research Institute, Durham, North Carolina 27701, USA \printead{e1}.}
\address[B]{Anastasios A. Tsiatis is Gertrude M. Cox Distinguished Professor,
Department
of Statistics, North Carolina State University, Raleigh, North
Carolina 27695-8203, USA \printead{e2}.}
\address[C]{Eric B. Laber is Assistant Professor, Department
of Statistics, North Carolina State University, Raleigh,
North Carolina 27695-8203, USA \printead{e3}.}
\address[D]{Marie Davidian is William Neal Reynolds Professor, Department
of Statistics, North Carolina State University, Raleigh, North Carolina
27695-8203, USA \printead{e4}.}
\end{aug}

%
\begin{abstract}
In clinical practice, physicians make a series of treatment
decisions over the course of a patient's disease based on his/her
baseline and evolving characteristics. A dynamic treatment regime
is a set of sequential decision rules that operationalizes this
process. Each rule corresponds to a decision point and dictates the
next treatment action based on the accrued information. Using
existing data, a key goal is estimating the optimal regime, that, if
followed by the patient population, would yield the most favorable
outcome on average. \textit{Q}- and \textit{A}-learning are two main
approaches for this purpose. We provide a detailed account of these
methods, study their performance, and illustrate them using data
from a depression study.
\end{abstract}

%

\end{frontmatter}
%
\section{Introduction}
\label{s:intro}

An area of current interest is personalized medi\-cine, which involves
making treatment decisions for an individual patient using all
information available on the patient, including genetic, physiologic,
demographic and other clinical variables, to achieve the ``best''
outcome for the patient given this information. In treating a patient
with an ongoing disease or disorder, a clinician makes a series of
decisions based on the patient's evolving status. A dynamic treatment
regime is a list of sequential decision rules formalizing this
process. Each rule corresponds to a key decision point in the
disease/disorder progression and takes as input the information on the
patient to that point and outputs the treatment that s/he should
receive from among the available options. A key step toward
personalized medicine is thus finding the optimal dynamic treatment
regime, that which, if followed by the entire patient population,
would yield the most favorable outcome on average.

The statistical problem is to estimate the optimal regime based on
data from a clinical trial or observational study. \iQ-learning
($Q$ denoting ``quality,'' \cite{Watkins,WatkinsDayan,Nahum}) and
advantage learning (\iA-learning, \cite{Murphy03,Robins04,Blatt})
are two main approaches for this purpose and are related to
reinforcement learning methods for sequential decision-making in
computer science. \iQ-learning is based roughly on posited regression
models for the outcome of interest given patient information at each
decision point and is implemented through a backward recursive
fitting procedure that is related to the dynamic programming algorithm
(\citep{Bather}), a standard approach for deducing optimal sequential
decisions. \iA-learning involves the same recursive strategy, but
requires only posited models for the part of the outcome regression
representing contrasts among treatments and for the probability of
observed treatment assignment given patient information at each
decision point. As discussed later, this may make \mbox{\iA-learning} more
robust to model misspecification than \iQ-learning for consistent
estimation of the optimal treatment regime.

Examples of the use of \iQ- and \iA-learning and alternative methods
to deduce optimal strategies for treatment of substance abuse,
psychiatric disorders, cancer and HIV infection and for dose
adjustment in response to evolving patient status have been presented
(\cite{Rosthoj}, \citeauthor{Murphy07a}, \citeyear{Murphy07a,Murphy07b},
\cite{Zhao,Henderson}).
Relevant work includes \citet{Thall00}, \citet{Thall02},
\citet{Robins04}, \citet{Demystifying}, \citet{Thall07}, \citet{van},
\citet{Orellana08}, \citet{Almirall}, \citet{Orellana10},
\citeauthor{Stat} (\citeyear{Stat,Zhang12,Zhang13})
and
\citeauthor{Zhao2} (\citeyear{Zhao2,Zhao3}).

The objective of this article is to provide readers interested in an
introduction to estimation of optimal dynamic
treatment regimes with a
self-contained, detailed description of an appropriate statistical
framework in which to define formally an optimal regime, of some of
the operational and philosophical considerations involved, and of \iQ-
and \iA-learning methods. Section~\ref{s:framework} introduces the
statistical framework, and Sections~\ref{s:defining} and
\ref{s:midstream} discuss the form of the optimal regime. We describe
and contrast \iQ- and \iA-learning in Section~\ref{s:methods} and
present systematic empirical studies of their relative performance and
the effects of misspecification of the postulated models involved in
Section~\ref{s:simulations}. The methods are demonstrated using data
from the Sequenced Treatment Alternatives to Relieve Depression
(STAR*D, \cite{Rush}) study in Section~\ref{s:application}.


\section{Framework and Assumptions}
\label{s:framework}

Consider the setting of $K$ prespecified, ordered decision
points, indexed by $k=1,\ldots,K$, which may be times or events in the
disease or disorder process that necessitate a treatment decision,
where, at each point, a~set of treatment options is available. Assume
that there is a final outcome $Y$ of interest for which large values
are preferred. The outcome may be ascertained following the $K$th
decision, as with CD4 T-cell count at a prespecified follow-up time in
HIV infection (\citep{Demystifying}), or may be a function of
information accrued over the entire sequence of decisions, as in
\citet{Henderson}, where the outcome is the overall proportion of time a
measure of blood clotting speed is kept within a target range in
dosing of anticoagulant agents.

In order to define an optimal treatment regime and discuss its
estimation based on data from an observational study or clinical
trial, we define a suitable conceptual framework. For simplicity, our
presentation is heuristic. Imagine that there is a superpopulation of
patients, denoted by $\Omega$, where one may view an element $\omega
\in\Omega$ as a patient from this population. We assume that patients
in the population have been treated according to routine clinical
practice for the disease or disorder prior to the first treatment
decision. Consequently, immediately prior to this first decision,
patient $\omega$ would present to the decision-maker with a set of
baseline information (covariates) denoted by the random variable
$S_1$, discussed further below. Thus, $S_1(\omega)$ is the value of
his/her information immediately prior to decision 1, taking values
$s_1$, say, in a set $\mS_1$. Assume that, at each decision point
$k=1,\ldots,K$, there is a finite set of all possible treatment
options $\mA_k$, with elements $a_k$. We do not consider the case of
continuous treatment and henceforth restrict attention to a finite set of
options. Denote by $\bar{a}_k = (a_1,\ldots,a_k)$ a possible treatment
history that could be administered through decision $k$, taking values
in $\barmA_k=\mA_1 \times\cdots\times\mA_k$,
where $\bar{\mathcal{A}}_K$ is
the set of all
possible treatment histories $\bar{a}_K$ through all $K$ decisions.

We then define the potential outcomes (\citep{Robins86})
%
\begin{eqnarray}\label{eq:potential}
W^*&=& \bigl\{S_2^*(a_1),S_3^*(
\bar{a}_2),\ldots,S_k^*(\bar{a}_{k-1}),\ldots,
\nonumber
\\[-8pt]
\\[-8pt]
\nonumber
&& \hspace*{ 6pt}S_K^*(\bar{a}_{K-1}),Y^*(\bar{a}_K)
\mbox{ for all } \bar{a}_K \in\barmA_K \bigr\}.
\end{eqnarray}
In (\ref{eq:potential}), $S^*_k(\bar{a}_{k-1})(\omega)$ denotes the
value of covariate information that would arise between decisions
$k-1$ and $k$ for a patient $\omega\in\Omega$ in the hypothetical
situation that s/he were to have previously received treatment history
$\bar{a}_{k-1}$, taking values $s_k$ in a set $\mS_k$,
$k=2,\ldots,K$. Similarly, $Y^*(\bar{a}_K)(\omega)$ is the hypothetical
outcome that would result for $\omega$ were s/he to have been
administered the full set of $K$ treatments in $\bar{a}_K$. This
notation implies that, for random variables such as
$S^*_k(\bar{a}_{k-1})$, $\bar{a}_{k-1}$ is an index representing prior
treatment history. Write $\barS^*_k(\bar{a}_{k-1}) =
\{S_1,S^*_2(a_1),\ldots,S^*_k(\bar{a}_{k-1})\}$, $k=1,\ldots,K$, where
$\barS^*_k(\bar{a}_{k-1})(\omega)$ takes values $\bar{s}_k$ in
$\barmS_k =
\mS_1 \times\cdots\times\mS_k$; this definition includes the
baseline covariate $S_1$ and is taken equal to $S_1$ when $k=1$. The
elements of the $\barS^*_k(\bar{a}_{k-1})$ and $Y^*(\bar{a}_K)$ may be
discrete or continuous; in what follows, for simplicity, we take these
random variables to be discrete, but the results hold more generally.

A dynamic treatment regime $d=(d_1,\ldots,d_K)$ is a set of rules that
forms an algorithm for treating a patient over time; it is ``dynamic''
because treatment is determined based on a patient's previous history.
At the $k$th decision point, the $k$th rule
$d_k(\bar{s}_k,\bar{a}_{k-1})$, say, takes as input the patient's realized
covariate and treatment history prior to the $k$th treatment decision
and outputs a value $a_k \in\Psi_k(\bar{s}_k,\bar{a}_{k-1})
\subseteq
\mA_k$; for $k=1$, there is no prior treatment ($a_0$ is null), and we
write $d_1(s_1)$ and $\Psi_1(s_1)$. Here,
$\Psi_k(\bar{s}_k,\bar{a}_{k-1})$ is a specified set of possible treatment
options for a patient with realized history $(\bar{s}_k,\bar{a}_{k-1})$,
discussed further below. Accordingly, although we suppress this in
the notation for brevity, the definition of a dynamic treatment regime
we now present depends on the specified $\Psi_k(\bar{s}_k,\bar{a}_{k-1})$,
$k=1,\ldots,K$. Because $d_k(\bar{s}_k,\bar{a}_{k-1}) \in
\Psi_k(\bar{s}_k,\bar{a}_{k-1}) \subseteq\mA_k$, $d_k$ need only
map a
subset of $\barmS_k \times\barmA_{k-1}$ to $\mA_k$. We define these
subsets recursively as
%
\begin{eqnarray}\label{eq:gammaconditions}
&& \Gamma_k=
\bigl\{(\bar{s}_k,\bar{a}_{k-1})\in\barmS_k
\times \barmA_{k-1} \mbox{ satisfying}\nonumber\\
&& \hspace*{31pt}\mbox{(i) } a_j \in
\Psi _j(\bar{s}_j,\bar{a}_{j-1}), j=1,
\ldots,k-1 \mbox{ and}\\
&& \hspace*{31pt}\mbox{(ii) } \pr \bigl\{\barS^*_k(
\bar{a}_{k-1})=\bar{s}_k \bigr\}>0 \bigr\},k=1, \ldots,K,
\nonumber
\end{eqnarray}
determined by $\Psi= (\Psi_1,\ldots,\Psi_K)$. The
$\Gamma_k$ contain all realizations of covariate and treatment history
consistent with having followed such
$\Psi$-specific regimes to decision $k$. Define the class $\mD$ of
($\Psi$-specific) dynamic treatment regimes to be the set of all
$d$ for which $d_k$, $k=1,\ldots,K$, is a mapping from
$\Gamma_k$ into $\mathcal{A}_k$ satisfying $d_k(\bar{s}_{k},
\bar{a}_{k-1}) \in\Psi_k(\bar{s}_{k}, \bar{a}_{k-1})$ for every
$(\bar{s}_{k}, \bar{a}_{k-1})\in\Gamma_k$.

Specification of the $\Psi_k(\bar{s}_k,\bar{a}_{k-1})$, $k=1,\ldots
,K$, is
dictated by the scientific setting and objectives. Some treatment
options may be unethical or impossible for patients with certain
histories, making it natural to restrict the set of possible options
for such patients. In the context of public health policy, the focus
may be on regimes involving only treatment options that are less
costly or widely available unless a patient's condition is especially
serious, as reflected in his/her covariate information. In what
follows, we assume that a particular fixed set $\Psi$ is specified,
and by an optimal regime we mean an optimal regime within the class
of corresponding $\Psi$-specific regimes.

An optimal regime should represent the ``best'' way to intervene to
treat patients in $\Omega$. To formalize, for any $d \in\mD$,
writing $\bard_k=(d_1,\ldots,d_k)$, $ k=1,\ldots,K$, $\bard_K=d$,
define the potential outcomes associa\-ted with $d$ as
$\{S_2^*(d_1),\ldots,S^*_k(\bard_{k-1}),\ldots,\break S_K^*(\bard
_{K-1}),Y^*(d)\}$
such that, for any $\omega\in\Omega$, with $S_1(\omega)=s_1$,
%
\begin{eqnarray}\label{eq:shorthand}
d_1(s_1)&=&u_1,\nonumber\\
S^*_2(d_1) (\omega) &=& S^*_2(u_1)
(\omega)=s_2, \nonumber\\
 d_2(\bar{s}_2,u_1)&=&u_2,\quad
\ldots,\nonumber\\
 d_{K-1}(\bar{s}_{K-1},\baru_{K-2})&=&u_{K-1},
\\
 S^*_K(\bard_{K-1}) (\omega) &=& S^*_K(
\baru_{K-1}) (\omega)=s_K,\nonumber\\
 d_K(
\bar{s}_K,\baru_{K-1})&=& u_K,\nonumber\\
 Y^*(d) (
\omega)&=&Y^*( \baru_K) (\omega)=y. \nonumber
\end{eqnarray}
The index $\bard_{k-1}$ emphasizes that $S^*_k(\bard_{k-1})(\omega)$
represents the covariate information that would arise between
decisions $k-1$ and $k$ were patient $\omega$ to receive the
treatments sequentially dictated by the first $k-1$ rules in $d$.
Similarly, $Y^*(d)(\omega)$ is the final outcome that $\omega$ would
experience if s/he were to receive the $K$ treatments dictated by $d$.

With these definitions, the expected outcome in the population if all
patients with initial state $S_1=s_1$ were to follow regime
$d$ is $\E\{Y^*(d)|S_1=s_1\}$. An optimal regime, $d^{\mathrm
{opt}}\in\mD$,
say, satisfies
%
\begin{eqnarray}\label{eq:optimalregime}
\E \bigl\{Y^*(d)|S_1=s_1 \bigr\} \leq\E \bigl\{Y^*
\bigl( d^{\mathrm{opt}}\bigr)|S_1=s_1 \bigr\}
\nonumber
\\[-8pt]
\\[-8pt]
\eqntext{\mbox{for
all } \,d \in\mD \mbox{ and all } s_1 \in\mS_1.}
\end{eqnarray}
Because (\ref{eq:optimalregime}) is true for any fixed $s_1$, in
fact, $\E\{Y^*(d)\} \leq\E\{Y^*(d^{{\mathrm{opt}}})\}$ for any
$d \in\mD$.
In Section~\ref{s:defining}, we give the form of $d^{\mathrm{opt}}$ satisfying
(\ref{eq:optimalregime}).

Alternative specifications of $\Psi$ may lead to different classes of
regimes across which the optimal regime may differ. We emphasize that
the definition (\ref{eq:optimalregime}) is predicated on the
particular set $\Psi$, and hence class $\mD$, of interest. In
principle, the class $\mD$ of interest is conceived based on
scientific or policy objectives without reference to data available
from a particular study.

Of course, potential outcomes for a given patient for all $d \in\mD$
are not observed. Thus, the goal is to estimate $d^{\mathrm{opt}}$ in
(\ref{eq:optimalregime}) using data from a study carried out on a
random sample of $n$ patients from $\Omega$ that record baseline and
evolving covariate information and treatments actually received.
Denote these available data as independent and identically
distributed (i.i.d.) time-ordered random variables
$(S_{1i},A_{1i},\ldots,S_{Ki},A_{Ki},Y_i)$, $i=1,\ldots,n$, on
$\Omega$.
Here, $S_1$ is as before; $S_k$, $k=2,\ldots,K$, is covariate
information recorded between decisions $k-1$ and $k$, taking values
$s_k \in\mS_k$; $A_k$, $k=1,\ldots, K$, is the recorded, observed
treatment assignment, taking values $a_k \in\mA_k$; and $Y$ is the
observed outcome, taking values $y \in\mY$. As above, define
$\barS_k = (S_1,\ldots,S_k)$ and $\barA_k=(A_1,\ldots,A_k)$,
$k=1,\ldots,K$, taking values $\bar{s}_k \in\barmS_k$ and $\bar
{a}_k \in
\barmA_k$.

The available data may arise from an observational study involving $n$
participants randomly sampled from the population; here,
treatment assignment takes place according to routine clinical
practice in the population. Alternatively, the data may arise from an
intervention study. A clinical trial design that has been advocated
for collecting data suitable for estimating optimal treatment regimes
is that of a so-called sequential multiple-assignment randomized trial
(SMART, \cite{Lavori,Murphy05}). In a SMART involving $K$
pre-specified decision points, each participant is randomized at each
decision point to one of a set of treatment options, where, at the
$k$th decision, the randomization probabilities may depend on past
realized information $\bar{s}_k,\bar{a}_{k-1}$.

In order to use the observed data from either type of study to
estimate an optimal regime, several assumptions are required. As is
standard, we make the consistency assumption (e.g., \cite{Robins94})
that the covariates and outcomes observed in the study are those that
potentially would be seen under the treatments actually received, that
is, $S_k = S_k^*(\barA_{k-1})$, $k=2,\ldots,K$, and $Y=Y^*(\barA_K)$.
We also make the stable unit treatment value assumption
(\citep{Rubin78}), which ensures that a patient's covariates and outcome
are unaffected by how treatments are allocated to her/him and other
patients. The critical assumption of no unmeasured confounders, also
referred to as the sequential randomization assumption
(\citep{Robins94}), must be satisfied. A strong version of this
assumption states that $A_k$ is conditionally independent of $W^*$ in
(\ref{eq:potential}) given $\{\barS_k,\barA_{k-1}\}$, $k=1,\ldots,K$,
where $A_0$ is null, written $A_k \independent W^* |
\barS_k,\barA_{k-1}$. In a SMART, this assumption is satisfied by
design; in an observational study, it is unverifiable from the
observed data. The strong version is sufficient for identification of the
distribution of not only $Y^*(\bar{a}_K)$ but of the joint distribution
of $Y^*(\bar{a}_K)$ and $\barS_K^*(\bar{a}_{K-1})$ and allows
the results of Section~\ref{s:midstream} to hold. Although in the
population patients and their providers may make decisions based only
on past covariate information available to them, the issue is whether
or not all of the information that is related to treatment assignment
and future covariates and outcome is recorded in the $S_k$; see
\citeauthor{Robins04} [(\citeyear{Robins04}), Sections~2--3] for discussion and a relaxation of the
version of the sequential randomization assumption given here. We
assume henceforth that these assumptions hold.

Whether or not it is possible to estimate $d^{\mathrm{opt}}$ from the available
data is predicated on the treatment options in
$\Psi_k(\bar{s}_k,\bar{a}_{k-1})$, $k=1,\ldots,K$, being
represented in
the data. For a prospectively-designed SMART, ordinarily, $\Psi$
defining the class $\mD$ of interest would dictate the design. At
decision $k$, subjects would be randomized to the options in
$\Psi_k(\bar{s}_k,\bar{a}_{k-1})$, satisfying this condition. If the data
are from an observational study, all treatment options in
$\Psi_k(\bar{s}_k,\bar{a}_{k-1})$ at each decision $k$ must have
been assigned
to some patients. That~is, if we define recursively $\Gamma^{\mathrm
{max}}_1 =
\{s_1 \in\mS_1\dvtx\pr(S_1=s_1)>0\}$, $\Psi^{\mathrm{max}}_1(s_1)
=\{a_1 \in
\mA
_1\dvtx\pr(A_1=a_1|S_1=s_1)>0$ for all $s_1 \in\Gamma
^{\mathrm{max}}_1\}$,
$\Gamma^{\mathrm{max}}_k=[(\bar{s}_k,\bar{a}_{k-1})\in\barmS_k
\times\barmA_{k-1}$ satisfying (i) $a_j \in
\Psi^{\mathrm{max}}_j(\bar{s}_j,\bar{a}_{j-1}), j=1,\ldots,  k-1$,
and  (ii) $\pr\{\barS^*_k(\bar{a}_{k-1})=\bar{s}_k\}>0 ]$,
$\Psi^{\mathrm{max}}_k(\bar{s}_k,  \bar{a}_{k-1}) = \{ a_k \in\mA
_k\dvtx\pr
(A_k=a_k|\barS_k=\bar{s}_k,\barA_{k-1}=  \bar{a}_{k-1})>0$ for all
$(\bar{s}_k,\bar{a}_{k-1}) \in\Gamma^{\mathrm{max}}_k\}$,
$k=2,\ldots,K$, we must
have $\Psi_k(\bar{s}_k,\bar{a}_{k-1}) \subseteq
\Psi^{\mathrm{max}}_k(\bar{s}_k,\bar{a}_{k-1})$, $k=1,\ldots,K$.
The class of regimes
dictated by $\Psi^{\mathrm{max}}=(\Psi^{\mathrm{max}}_1,\ldots
,\Psi^{\mathrm{max}}_K)$ is the largest that
can be considered based on the data, sometimes referred to as the
class of ``feasible regimes'' (\citep{Robins04}). If this inclusion
condition does not hold
for all $k=1,\ldots,K$, $d^{\mathrm{opt}}$ cannot be estimated from
the data, and
the class of regimes $\mD$ of interest must be reevaluated or another
data source found.

\section{Optimal Treatment Regimes}
\label{s:defining}

\iQ- and \iA-learning are two approaches to estimating $d^{\mathrm{opt}}$
satisfying (\ref{eq:optimalregime}) under the foregoing framework.
Both involve recursive fitting algorithms; the main distinguishing
feature is the form of the underlying models. To appreciate the
rationale, one must understand how $d^{\mathrm{opt}}$ is determined
via dynamic
programming, also known as backward induction. We demonstrate the
formulation of $d^{\mathrm{opt}}$ in terms of the potential outcomes
and then
show how $d^{\mathrm{opt}}$ may be expressed in terms of the observed
data under
assumptions including those in Section~\ref{s:framework}. We sometimes
highlight dependence on specific elements of quantities such as
$\bar{a}_k$, writing, for example, $\bar{a}_k$ as $(\bar{a}_{k-1},a_k)$.

At the $K$th decision point, for any $\bar{s}_K \in\barmS_K$,
$\bar{a}_{K-1} \in\barmA_{K-1}$ for which
$(\bar{s}_K,\bar{a}_{K-1}) \in\Gamma_K$, define
%
\begin{eqnarray}
\label{eq:doptK}
&&d^{(1){\mathrm{opt}}}_K(\bar{s}_K,\bar{a}_{K-1})
\nonumber
\\
&&\quad= \operatorname{arg}\max_{a_K \in
\Psi_K(\bar{s}_K,\bar{a}_{K-1})} \E \bigl\{Y^*(\bar{a}_{K-1},a_K)
| \\
&&\hspace*{118pt}\barS^*_K(\bar{a}_{K-1})= \bar{s}_K \bigr\},\nonumber
\\
\label{eq:VK}
&&V^{(1)}_K(\bar{s}_K,\bar{a}_{K-1})
\nonumber
\\
&&\quad
=  \max_{a_K \in
\Psi_K(\bar{s}_K,\bar{a}_{K-1})} \E \bigl\{Y^*(\bar{a}_{K-1},a_K)
|\\
&& \hspace*{102pt}\barS^*_K(\bar{a}_{K-1})= \bar{s}_K \bigr\}.\nonumber
\end{eqnarray}
%
For $k=K - 1,\ldots,1$ and any $\bar{s}_k \in\barmS_k$, $\bar{a}_{k-1}
\in\barmA_{k-1}$ for which $(\bar{s}_k,\bar{a}_{k-1}) \in\Gamma_k$,
which clearly holds if $(\bar{s}_K,\bar{a}_{K-1}) \in\Gamma_K$, let
%
\begin{eqnarray}\label{eq:doptk}
&&d^{(1){\mathrm{opt}}}_k(\bar{s}_k,
\bar{a}_{k-1}) \nonumber\\
&&\quad =  \operatorname{arg} \max_{a_k \in
\Psi_k(\bar{s}_k,\bar{a}_{k-1})} \E \bigl[
V^{(1)}_{k+1} \bigl\{\bar{s}_k,S^*_{k+1}(
\bar{a}_{k-1},a_k),
\\
&&\hspace*{115pt}\bar{a}_{k-1},a_k
\bigr\} |\barS^*_k(\bar{a}_{k-1})=\bar{s}_k
\bigr], \nonumber
\\
\label{eq:Vk}
&&V^{(1)}_k(\bar{s}_k,\bar{a}_{k-1})\nonumber\\
&&\quad
= \max_{a_k \in\Psi_k(\bar{s}_k,\bar{a}_{k-1})} \E \bigl[ V^{(1)} _{k+1} \bigl
\{\bar{s}_k,S^*_{k+1}( \bar{a}_{k-1},a_k),
\\
\nonumber
&&\hspace*{115pt}
\bar{a}_{k-1},a_k \bigr\} | \barS^*_k(
\bar{a}_{k-1})=\bar{s}_k \bigr]; \nonumber
\end{eqnarray}
thus, for $s_1 \in\mS_1$,
\begin{eqnarray*}
&&d^{(1){\mathrm{opt}}}_1(s_1) \\
&&\quad= \operatorname{arg} \max_{a_1
\in\Psi_1(s_1)} \E\bigl[ V^{(1)}_2\bigl\{s_1,S^*_2(a_1),a_1\bigr\}|S_1=s_1\bigr],
\\
&&V^{(1)}_1(s_1) \\
&&\qquad= \max_{a_1 \in\Psi_1(s_1)} \E\bigl[
V^{(1)}_2\bigl\{s_1,S^*_2(a_1),a_1\bigr\}|S_1=s_1\bigr].
\end{eqnarray*}
Conditional expectations
are well defined by (\ref{eq:gammaconditions})(ii).

Clearly, $d^{(1){\mathrm{opt}}}=(d^{(1){\mathrm{opt}}}_1,\ldots
,d^{(1){\mathrm{opt}}}_K)$ is a treatment
regime, as it comprises a set of rules that uses patient information
to assign treatment from among the options in $\Psi$. The superscript
(1) indicates that $d^{(1){\mathrm{opt}}}$ provides $K$ rules for a patient
presenting prior to decision 1 with baseline information $S_1=s_1$;
Section~\ref{s:midstream} considers optimal treatment of patients presenting
at subsequent decisions after receiving possibly suboptimal treatment
at prior decisions. Note that $d^{(1){\mathrm{opt}}}$ is defined in a backward
iterative fashion. At decision $K$, (\ref{eq:doptK}) gives the
treatment that maximizes the expected potential final outcome given
the prior potential information, and (\ref{eq:VK}) is the maximum
achieved. At decisions $k=K-1,\ldots,1$, (\ref{eq:doptk}) gives the
treatment that maximizes the expected outcome that would be achieved
if subsequent optimal rules already defined were followed
henceforth. In Section~A.1 of the supplemental article
[\citet{Schulte}], we show that $d^{(1){\mathrm{opt}}}$ defined in
(\ref{eq:doptK})--(\ref{eq:Vk}) is an optimal treatment regime in the
sense of satisfying (\ref{eq:optimalregime}).

The foregoing developments express optimal  regi\-mes in terms of the
distribution of potential outcomes. If an optimal regi\-me is to be
identifiable, it must be possible under the assumptions in
Section~\ref{s:framework} to express $d^{(1){\mathrm{opt}}}$ in terms
of the
distribution of the observed data. To this end, define
%
\begin{eqnarray}
&&Q_K(\bar{s}_K,\bar{a}_K) =  \E(Y|
\barS_K=\bar{s}_K,\barA_K=
\bar{a}_K), \label{eq:qK}
\\
 \label{eq:dfinaloptK}
 &&d^{\mathrm{opt}}_K(\bar{s}_K,\bar{a}_{K-1})
\nonumber
\\[-8pt]
\\[-8pt]
\nonumber
&&\quad
=  \operatorname{arg} \max_{a_K \in
\Psi_K(\bar{s}_K,\bar{a}_{K-1})} Q_K(
\bar{s}_K, \bar{a}_{K-1},a_K),
\\
 \label{eq:VfinalK}
 &&V_K(\bar{s}_K,\bar{a}_{K-1})
\nonumber
\\[-8pt]
\\[-8pt]
\nonumber
&&\quad  =  \max
_{a_K \in
\Psi_K(\bar{s}_K,\bar{a}_{K-1})} Q_K(\bar{s}_K,
\bar{a}_{K-1},a_K),
\end{eqnarray}
and for $k=K-1,\ldots,1$, define
%
\begin{eqnarray}
\label{eq:qk}
&& Q_k(\bar{s}_k,\bar{a}_k)
\nonumber
\\[-8pt]
\\[-8pt]
\nonumber
&&\quad =  \E \bigl\{
V_{k+1}(\bar{s}_k,S_{k+1},\bar{a}_k)
| \barS_k=\bar{s}_k,\barA_k=
\bar{a}_k \bigr\},
\\
\label{eq:dfinaloptk}
\quad &&d^{\mathrm{opt}}_k(\bar{s}_k,\bar{a}_{k-1})
\nonumber
\\[-8pt]
\\[-8pt]
\nonumber
&&\quad
=  \operatorname{arg} \max_{a_k \in
\Psi_k(\bar{s}_k,\bar{a}_{k-1})} Q_k(
\bar{s}_k, \bar{a}_{k-1},a_k),
\\
&&V_k(\bar{s}_k,\bar{a}_{k-1})  =  \max
_{a_k \in
\Psi_k(\bar{s}_k,\bar{a}_{k-1})} Q_k(\bar{s}_k,
\bar{a}_{k-1},a_k). \label{eq:Vfinalk}
\end{eqnarray}
The expressions in (\ref{eq:qK})--(\ref{eq:Vfinalk}) are well defined
under assumptions we discuss next. In (\ref{eq:qK}) and
(\ref{eq:qk}), $Q_k(\bar{s}_k,\bar{a}_k)$ are referred to as
``\iQ-functions,'' viewed as measuring the ``quality'' associated with
using treatment $a_k$ at decision $k$ given the history up to that
decision and then following the optimal regime thereafter. The ``value
functions'' $V_k(\bar{s}_k,\bar{a}_{k-1})$ in (\ref{eq:VfinalK}) and
(\ref{eq:Vfinalk}) reflect the ``value'' of a patient's history
$\bar{s}_k,\bar{a}_{k-1}$ assuming that optimal decisions are made in the
future. We emphasize that the $d^{\mathrm{opt}}_k$, $k=1,\ldots,K$,
defined in
(\ref{eq:qK})--(\ref{eq:Vfinalk}) may not be optimal unless the
sequential randomization, consistency and positivity assumptions
hold.

As in Section~\ref{s:framework}, the treatment options in
$\Psi$ must be represented in the data, that is,
$\Psi_k(\bar{s}_k,\bar{a}_{k-1}) \subseteq
\Psi^{\mathrm{max}}_k(\bar{s}_k,\bar{a}_{k-1})$, $k=1,\ldots,K$,
in order to estimate
an optimal regime. Formally, this implies that
%
\begin{eqnarray}\label{eq:positivity}
\pr(A_k=a_k|\barS_k=\bar{s}_k,
\barA_{k-1}=\bar{a}_{k-1})> 0
\nonumber
\\[-8pt]
\\[-8pt]
\eqntext{ \mbox{if } (
\bar{s}_k, \bar{a}_{k-1}) \in\Gamma_k \mbox{
and } a_k \in \Psi_k(\bar{s}_k,
\bar{a}_{k-1})}
\end{eqnarray}
for all $k=1,\ldots,K$. In Section~A.2 of the supplemental article
[\citet{Schulte}],
under the consistency and sequential randomization assumptions
and the positivity assumption (\ref{eq:positivity}), we show that,
for any $(\bar{s}_k,\bar{a}_{k-1})
\in\Gamma_k$ and $a_k \in\Psi_k(\bar{s}_k,\bar{a}_{k-1})$,
$k=1,\ldots,K$,
%
\begin{eqnarray}\quad\hspace*{8pt}
&&\pr(\barS_k=\bar{s}_k,\barA_k=
\bar{a}_k) > 0, \label{eq:ma0}
\\
\label{eq:ma1}
 &&\pr(S_{k+1}=s_{k+1} |
\barS_k=\bar{s}_k,\barA_k=
\bar{a}_k)
\nonumber
\\[-8pt]
\\[-8pt]
\nonumber
&&\quad= \pr \bigl\{S^*_{k+1}(\bar{a}_k)=s_{k+1}
| \barS_k=\bar{s}_k,\barA_{k-1}=\bar{a}
_{k-1} \bigr\}
\\
\label{eq:ma2}
&&\quad = \pr \bigl\{S^*_{k+1}(\bar{a}_k)=s_{k+1} |
\barS_j=\bar{s}_j, \barA_{j-1}=
\bar{a}_{j-1},
\nonumber
\\[-8pt]
\\[-8pt]
\nonumber
&&\hspace*{50pt}S^*_{j+1}(\bar{a}_j)=s_{j+1},
\ldots, S^*_k(\bar{a}_{k-1})=s_k \bigr\},
\end{eqnarray}
for $j=1,\ldots,k$, where (\ref{eq:ma2}) with $j=k$ is the same as the
right-hand side of (\ref{eq:ma1}), $S_{K+1}=Y$ and
$S^*_{K+1}(\bar{a}_K)=Y^*(\bar{a}_K)$, and when $j=1$ the conditioning
events do not involve treatment. By (\ref{eq:ma0}), the quantities in
(\ref{eq:qK})--(\ref{eq:Vfinalk}) are well defined. Under
(\ref{eq:ma1})--(\ref{eq:ma2}), the conditional distributions of the
observed data involved in (\ref{eq:qK})--(\ref{eq:Vfinalk}) are the
same as the conditional distributions of the potential outcomes
involved in (\ref{eq:doptK})--(\ref{eq:Vk}). It follows that
%
\begin{eqnarray} \label{eq:dequiv1}
d^{(1){\mathrm{opt}}}_k(\bar{s}_k,\bar{a}_{k-1}) &=&
d^{\mathrm{opt}}_k(\bar{s}_k,\bar{a}_{k-1}),
\nonumber
\\[-8pt]
\\[-8pt]
\nonumber
V^{(1)}_k(\bar{s}_k,\bar{a}_{k-1}) &=&
V_k(\bar{s}_k,\bar{a}_{k-1}),
\end{eqnarray}
for $(\bar{s}_k,\bar{a}_{k-1}) \in\Gamma_k$, $k=1,\ldots,K$. The
equivalence in (\ref{eq:dequiv1}) shows that, under the consistency,
sequential randomization and positivity assumptions, an optimal
treatment regime in the ($\Psi$-specific) class of interest $\mD$ may
be obtained using the distribution of the observed data.

There may not be a unique $d^{\mathrm{opt}}$. At any decision $k$, if
there is
more than one possible option $a_k$ maximizing the \iQ-function,
then any rule $d^{\mathrm{opt}}_k$ yielding one of these $a_k$ defines
an optimal
regime.

\section{Optimal ``Midstream'' Treatment Regime}
\label{s:midstream}

In Section~\ref{s:defining} we define an ($\Psi$-specific) optimal
treatment regime starting at decision point 1 and elucidate conditions
under which it may be estimated using data from a clinical or
observational study collected through all $K$ decisions on a sample
from the patient population. The goal is to estimate the optimal
regime and implement it in new such patients presenting at the first
decision.

In routine clinical practice, however, a new patient may be
encountered subsequent to decision point 1. For definiteness, suppose
a new patient presents ``midstream,'' immediately prior to the
$\ell$th decision point, $\ell=2,\ldots,K$. A natural question is how
to treat this patient optimally henceforth. For such a patient, the
first $\ell-1$ treatment decisions presumably have been made according
to routine practice, and s/he has a realized past history that may be
viewed as realizations of random variables
$(S^{(P)}_1,A^{(P)}_1,\ldots,S^{(P)}_{\ell-1},A^{(P)}_{\ell
-1},S^{(P)}_\ell)$. Here,
$A^{(P)}_k$, $k=1,\ldots, \ell-1$, represent the treatments received by
such a patient according to the treatment assignment mechanism
governing routine practice; and $S^{(P)}_k$, $k=1,\ldots, \ell-1$, denote
covariate information collected up to the $\ell$th decision. Write
$\bar{A}^{(P)}_k = (A^{(P)}_1,\ldots,A^{(P)}_k)$, $k=1,\ldots,\ell
-1$, and
$\bar{S}^{(P)}_k=(S^{(P)}_1,\ldots,S^{(P)}_k)$, $k=1,\ldots,\ell$.

As $\mA_k$ denotes the set of all possible treatment options at
decision $k$, $\bar{A}^{(P)}_{\ell-1}$ takes on values $\bar
{a}_{\ell-1}
\in
\barmA_{\ell-1}$. To define $\Psi$-specific regimes starting at
decision $\ell$, at the least, $S^{(P)}_k$ must contain the same
information as $S_k$ in the data, $k=1,\ldots,\ell$. Because the
available data dictate the covariate information incorporated in the
class of regimes $\mD$, if $S^{(P)}_k$ contains additional
information, it
cannot be used in the context of such regimes. We thus take $S^{(P)}_k$
and $S_k$ to contain the same information, stated formally as the
consistency assumption $S^{(P)}_k=S^*_k(\bar{A}^{(P)}_{k-1})$,
$k=1,\ldots,\ell$.
Moreover, we can only consider treating new patients with realized
histories $(\bar{s}_\ell,\bar{a}_{\ell-1})$ that are contained in
$\Gamma_\ell$, that is, that could have resulted from following a
$\Psi$-specific regime through decision $\ell-1$. If the data arise
from a SMART including only a subset of the treatments employed in
practice, this may not hold.

We thus desire rules $d^{(\ell)}_k(\bar{s}_k,\bar{a}_{k-1})$,
$k=\ell,\ell+1,\ldots,K$, say, that dictate how to treat such
midstream patients presenting with realized past history
$(\bar{S}^{(P)}_\ell,\bar{A}^{(P)}_{\ell-1}) = (\bar{s}_\ell,\bar
{a}_{\ell
-1})$. In the
following, we regard $(\bar{s}_\ell,\bar{a}_{\ell-1})$ as fixed,
corresponding to the particular new patient. Let $\Gamma^{(\ell)}_k$
be all
elements of $\Gamma_k$ with $(\bar{s}_\ell,\bar{a}_{\ell-1})$
fixed at the
values for the given new patient. Write
$d^{(\ell)}=(d^{(\ell)}_\ell,d^{(\ell)}_{\ell+1},\ldots,d^{(\ell
)}_K)$ to denote regimes
starting at the $\ell$th decision point, and define the class
$\mD^{(\ell)}$ of all such regimes to be the set of all $d^{(\ell)}$ for
which $d^{(\ell)}_k(\bar{s}_k,\bar{a}_{k-1})=a_k$ for $(\bar
{s}_k,\bar{a}_{k-1})
\in\Gamma^{(\ell)}_k$ and $a_k \in\Psi_k(\bar{s}_k,\bar
{a}_{k-1})$ for
$k=\ell,\ldots,K$. Then, by analogy to (\ref{eq:optimalregime}), we
seek $d^{(\ell){\mathrm{opt}}}$ satisfying
%
\begin{eqnarray}\label{eq:optimalregimeell}\quad
&&\E \bigl\{Y^*\bigl(\bar{a}_{\ell-1},d^{(\ell)}\bigr) |
\bar{S}^{(P)}_\ell= \bar{s}_\ell,
\bar{A}^{(P)}_{\ell-1}=\bar{a}_{\ell-1} \bigr\}
\nonumber
\\[-8pt]
\\[-9pt]
\nonumber
&&\quad\leq \E \bigl
\{Y^*\bigl(\bar{a}_{\ell-1},d^{(\ell){\mathrm{opt}}}\bigr) | \bar
{S}^{(P)}_\ell=\bar{s}_\ell, \bar{A}^{(P)}_{\ell-1}=
\bar{a}_{\ell-1} \bigr\}
\end{eqnarray}
for all $d^{(\ell)}\in\mD^{(\ell)}$ and $\bar{s}_\ell\in\barmS
_\ell$,
$\bar{a}_{\ell-1} \in\barmA_{\ell-1}$ for which
$\pr(\bar{S}^{(P)}_\ell=\bar{s}_\ell,\bar{A}^{(P)}_{\ell-1}=\bar
{a}_{\ell-1})>0$.
Viewing this as a problem of making $K-\ell+1$ decisions at decision
points $\ell,\ell+1,\ldots,K$, with initial state
$\bar{S}^{(P)}_\ell=\bar{s}_\ell,\bar{A}^{(P)}_{\ell-1}=\bar
{a}_{\ell-1}$, by an
argument analogous to that in Section~A.1 of the supplemental article
[\citet{Schulte}] for
$\ell=1$ and initial state $S_1=s_1$, letting
$\mathcal{V}_{\ell,k}=\{\bar{S}^{(P)}_\ell=\bar{s}_\ell,\bar
{A}^{(P)}_{\ell
-1}=\bar{a}
_{\ell-1},S^*_{\ell+1}(\bar{a}_\ell)=
s_{\ell+1},\ldots,S^*_k(\bar{a}_{k-1})=s_k\}$, it may be shown that
$d^{(\ell){\mathrm{opt}}}$
satisfying (\ref{eq:optimalregimeell}) is given by
%
\begin{eqnarray}\label{eq:doptellK}
\quad&& d^{(\ell){\mathrm{opt}}}_K(\bar{s}_K,\bar{a}_{K-1})
\nonumber
\\[-8pt]
\\[-9pt]
\nonumber
&&\quad
=  \operatorname{arg} \max_{a_K \in
\Psi_K(\bar{s}_K,\bar{a}_{K-1})} \E \bigl\{Y^*(\bar{a}_{K-1},a_K)
| \mathcal{V}_{\ell,K} \bigr\},
\\
\label{eq:VellK}
&&V^{(\ell)}_K(\bar{s}_K,\bar{a}_{K-1})
\nonumber
\\[-8pt]
\\[-9pt]
\nonumber
&&\quad=  \max_{a_K \in
\Psi_K(\bar{s}_K,\bar{a}_{K-1})} \E \bigl\{Y^*(\bar{a}_{K-1},a_K)
| \mathcal{V}_{\ell
,K} \bigr\}
\end{eqnarray}
for any $\bar{s}_K \in\barmS_K$, $\bar{a}_{K-1} \in\barmA_{K-1}$ for
which $(\bar{s}_K,\break  \bar{a}_{K-1}) \in\Gamma^{(\ell)}_K$; and, for
$k=K-1,\ldots,\ell$,
%
\begin{eqnarray}
 \label{eq:doptellk}
&&\hspace*{-60pt}d^{(\ell){\mathrm{opt}}}_k(\bar{s}_k,\bar{a}_{k-1})\nonumber\\
&&\hspace*{-60pt}\quad
= \operatorname{arg} \max_{a_k \in
\Psi_k(\bar{s}_k,\bar{a}_{k-1})} \E \bigl[ 
V^{(\ell)}_{k+1} \bigl\{\bar{s}_k,
\\
\eqntext{{S^*_{k+1}(
\bar{a}_{k-1},a_k), \bar{a}_{k-1},a_k
\bigr\}  | 
\mathcal{V}_{\ell,k} \bigr],}}
\\
\label{eq:Vellk}
&&\hspace*{-60pt} V^{(\ell)}_k(\bar{s}_k,\bar{a}_{k-1})\nonumber\\
&&\hspace*{-60pt}\quad
= \max_{a_k \in
\Psi_k(\bar{s}_k,\bar{a}_{k-1})} \E \bigl[ 
V^{(\ell)}_{k+1} \bigl\{ \bar{s}_k,
\\
\eqntext{{S^*_{k+1}(
\bar{a}_{k-1},a_k), \bar{a}_{k-1},a_k
\bigr\} 
|\mathcal{V}_{\ell,k} \bigr] }}
\end{eqnarray}
for any $\bar{s}_k \in\barmS_k$, $\bar{a}_{k-1} \in\barmA_{k-1}$ for
which $(\bar{s}_k,\bar{a}_{k-1}) \in\Gamma^{(\ell)}_k$, so that
\begin{eqnarray*}
&&d^{(\ell){\mathrm{opt}}}_\ell(\bar{s}_\ell,\bar{a}_{\ell-1}) \\
&&\qquad=
\operatorname{arg} \max_{a_\ell\in\Psi_\ell(\bar{s}_\ell,\bar
{a}_{\ell-1})} \E \bigl[ 
V^{(\ell)}_{\ell+1} \bigl\{\bar{s}_\ell,S^*_{\ell+1}(
\bar{a}_{\ell-1},a_\ell),\\ 
&&\hspace*{58pt}\qquad\bar{a}_{\ell-1},a_\ell
\bigr\}| \bar{S}^{(P)}_\ell=\bar{s}_\ell,
\bar{A}^{(P)}_{\ell-1}= \bar{a}_{\ell-1} \bigr].
\end{eqnarray*}

Comparison of (\ref{eq:doptK})--(\ref{eq:Vk}) to
(\ref{eq:doptellK})--(\ref{eq:Vellk}) shows that the $\ell$th to $K$th
rules of the optimal regime $d^{(1){\mathrm{opt}}}$ that would be
followed by a
patient presenting at the first decision are not necessarily the same
as those of the optimal regime $d^{(\ell){\mathrm{opt}}}$ that would
be followed by a
patient presenting at the $\ell$th decision. In particular, noting
that the conditioning sets in (\ref{eq:doptK})--(\ref{eq:Vk}) are
$\mathcal{V}_{1,K}$ and $\mathcal{V}_{1,k}$, the rules are
$\ell$-dependent through dependence of the conditioning sets
$\mathcal{V}_{\ell,k}$, $\ell=1,\ldots,K$, $k=\ell,\ldots,K$, on
$\ell$. However, we now demonstrate that these rules coincide under
certain conditions.

Make the consistency, sequential randomization and positivity
(\ref{eq:positivity}) assumptions on the available data required to
show (\ref{eq:dequiv1}) in
Section~\ref{s:defining}, along with the consistency assumption on the
$S^{(P)}_k$ above and the sequential randomization assumption $A^{(P)}_k
\independent W^* |\break \bar{S}^{(P)}_k,\bar{A}^{(P)}_{k-1}$, $k=1,\ldots
,\ell-1$, which
ensures that the $\bar{S}^{(P)}_k$ include all information related to
treatment assignment and future covariates and outcome up to decision
$\ell$. Note that (\ref{eq:doptellK})--(\ref{eq:Vellk}) are expressed
in terms of the conditional distributions
$\pr\{S^*_{k+1}(\bar{a}_k)=s_{k+1} | \bar{S}^{(P)}_\ell=\bar
{s}_\ell,
\bar{A}^{(P)}_{\ell-1}=\bar{a}_{\ell-1},S^*_{\ell+1}(\bar{a}_\ell
)=s_\ell,\ldots,\break
S^*_k(\bar{a}_{k-1})=s_k\}$, $k=\ell,\ldots,K$. We can then use
(\ref{eq:ma2}) with $j=\ell$ to deduce that these conditional
distributions can be written equivalently as
$\pr\{S^*_{k+1}(\bar{a}_k)=s_{k+1}|
\barS^*_k(\bar{a}_{k-1})=\bar{s}_k\}$, $k=\ell,\ldots,K$, so
solely in
terms of the distribution of the potential outcomes. By
(\ref{eq:ma1}) and (\ref{eq:ma2}) with $j=1$, this can be written as
$\pr(S_{k+1}=s_{k+1} | \barS_k = \bar{s}_k,\barA_k=\bar{a}_k)$. This
shows that (\ref{eq:doptellK})--(\ref{eq:Vellk}) can be reexpressed in
terms of the observed data, so that, for $(\bar{s}_k,\bar{a}_{k-1})
\in
\Gamma_k$ for $\ell=1,\ldots,K$ and $k=\ell,\ldots,K$,
%
\begin{eqnarray}\label{eq:dequiv}
d^{(\ell){\mathrm{opt}}}_k(\bar{s}_k,\bar{a}_{k-1}) &=&
d^{\mathrm{opt}}_k(\bar{s}_k,\bar{a}_{k-1}),
\nonumber
\\[-8pt]
\\[-8pt]
\nonumber
V^{(\ell)}_k(\bar{s}_k,\bar{a}_{k-1}) &=&
V_k(\bar{s}_k,\bar{a}_{k-1}).
\end{eqnarray}
Note that (\ref{eq:dequiv}) subsumes
(\ref{eq:dequiv1}) when $\ell=1$. The equivalence
in (\ref{eq:dequiv}) demonstrates not only that an
optimal treatment regime can be obtained using the distribution of the
observed data but also that the corresponding rules dictating
treatment do not depend on $\ell$ under these assumptions. Thus, the
single set of rules $d^{\mathrm{opt}}=(d^{\mathrm{opt}}_1,\ldots
,d^{\mathrm{opt}}_K)$ defined in
(\ref{eq:dfinaloptK}) and (\ref{eq:dfinaloptk}) is relevant regardless
of when a patient presents. That is, treatment at the $\ell$th
decision point for a patient who presents at decision 1 and has
followed the rules in $d^{\mathrm{opt}}$ to that point would be
determined by
$d^{\mathrm{opt}}_\ell$ evaluated at his/her history up to that
point, as would
treatment for a subject presenting for the first time immediately
prior to decision $\ell$. See \citeauthor{Robins04}
[(\citeyear{Robins04}), pages 305--306] for more
discussion.

\section{\textit{Q}- and \textit{A}-Learning}
\label{s:methods}
\subsection{Q-Learning}
\label{ss:qlearn}

From (\ref{eq:dfinaloptK}), (\ref{eq:dfinaloptk}) and
(\ref{eq:dequiv1}), an optimal ($\Psi$-specific) regime $d^{\mathrm
{opt}}$ may be
represented in terms of the \iQ-functions (\ref{eq:qK}),
(\ref{eq:qk}). Thus, estimation of $d^{\mathrm{opt}}$ based on i.i.d. data
$(S_{1i},A_{1i},\ldots,S_{Ki},A_{Ki},Y_i)$, $i=1,\ldots,n$, may be
accomplished via direct modeling and fitting of the \iQ-functions.
This is the approach underlying \iQ-learning. Specifically, one may
posit models $Q_k(\bar{s}_k,\bar{a}_k; \xi_k)$, say, for
$k=K,K-1,\ldots,1$, each depending on a finite-dimensional parameter
$\xi_k$. The models may be linear or nonlinear in $\xi_k$ and include
main effects and interactions in the elements of $\bar{s}_k$ and~$\bar{a}_k$.

Estimators $\hatxi_k$ may be obtained in a backward iterative fashion for
$k=K,K-1,\ldots,1$ by solving suitable estimating equations [e.g., ordinary
(OLS) or weighted (WLS) least squares]. Assuming the latter, for
$k=K$, letting $\tilV_{(K+1)i} = Y_i$, one would first solve
%
\begin{eqnarray}\label{eq:qlearnK}
&&\sum_{i=1}^n \frac{\partial Q_K(\barS_{Ki},\barA_{Ki};\xi
_K)}{\partial
\xi_K}
\Sigma_K^{-1}(\barS_{Ki},\barA_{Ki})
\nonumber
\\[-8pt]
\\[-8pt]
\nonumber
&&\quad{}\times
\bigl\{\tilV_{(K+1)i} - Q_K(\barS_{Ki},
\barA_{Ki}; \xi_K) \bigr\}=0
\end{eqnarray}
in $\xi_K$ to obtain $\hatxi_K$, where $\Sigma_K(\bar{s}_K,\bar
{a}_K)$ is
a working variance model. Substituting the model
$Q_K(\bar{s}_K,\bar{a}_K;\break   \xi_K)$ in (\ref{eq:dfinaloptK}) and accordingly
writing $d^{\mathrm{opt}}_K(\bar{s}_K,\bar{a}_{K-1};\break   \xi_K)$, substituting
$\hatxi_K$
for $\xi_K$, yields an estimator for the optimal treatment choice at
decision $K$ for a patient with past history
$\barS_K=\bar{s}_K,\barA_{K-1}=\bar{a}_{K-1}$. With $\hatxi_K$ in hand,
one would form for each $i$, based on (\ref{eq:VfinalK}), $\tilV_{Ki}
= \max_{a_K \in\Psi_K(\barS_{Ki},\barA_{(K-1)i})}
Q_K(\barS_{Ki},\break \barA_{(K-1)i},a_K; \hatxi_K)$. To obtain
$\hatxi_{K-1}$, setting $k=K-1$, based on (\ref{eq:qk}) and
letting $\Sigma_k(\bar{s}_k,\bar{a}_k)$ be a working variance model,
one would
then solve for $\xi_k$,
%
\begin{eqnarray}
\label{eq:qlearnk} &&\sum_{i=1}^n
\frac{\partial Q_k(\barS_{ki},\barA_{ki};\xi_k)}{\partial
\xi_k} \Sigma_k^{-1}(\barS_{ki},
\barA_{ki})
\nonumber
\\[-8pt]
\\[-8pt]
\nonumber
&&\quad{}\times  \bigl\{\widetilde{V}_{(k+1)i} - Q_k(
\barS_{ki},\barA_{ki}; \xi_k) \bigr\}=0.
\end{eqnarray}
The corresponding $d^{\mathrm{opt}}_{K-1}(\bar{s}_{K-1},\bar
{a}_{K-2}; \hatxi_{K-1})$
yields an estimator for the optimal treatment choice at decision
$K-1$ for a patient with past history
$\barS_{K-1}=\bar{s}_{K-1},\barA_{K-2}=\bar{a}_{K-2}$, assuming
s/he will
take the optimal treatment at decision $K$. One would continue this
process in the obvious fashion for $k=K-2,\ldots,1$, forming
$\tilV_{ki} = \max_{a_k \in\Psi_k(\barS_{ki},\barA_{(k-1)i})}
Q_k(\barS_{ki},\break \barA_{(k-1)i},  a_k; \hatxi_k)$, and solving equations
of form (\ref{eq:qlearnk}) to obtain $\hatxi_k$ and corresponding
$d^{\mathrm{opt}}_k(\bar{s}_k,\bar{a}_{k-1}; \hatxi_k)$.

We may now summarize the estimated optimal regime as $\widehat
{d}^{\mathrm{opt}}_Q =
(\widehat{d}^{\mathrm{opt}}_{Q,1},\ldots,\widehat{d}^{\mathrm
{opt}}_{Q,K})$, where
%
\begin{eqnarray}\label{eq:qestdopt}
\widehat{d}^{\mathrm{opt}}_{Q,1}(s_1) &=&
d^{\mathrm{opt}}_1(s_1; \hatxi_1),\nonumber\\
\widehat{d}^{\mathrm{opt}}_{Q,k}(\bar{s}_k,
\bar{a}_{k-1}) &=& d^{\mathrm{opt}}_k(\bar{s}_k,
\bar{a}_{k-1}; \hatxi_k),
\\
\eqntext{k=2,\ldots,K.}
\end{eqnarray}
It is important to recognize that, even under the sequential
randomization assumption, the estimated regime~(\ref{eq:qestdopt}) may
not be a consistent estimator for the true optimal regime unless all
the models for the \iQ-functions are correctly specified.

We illustrate the approach for $K=2$, where at each decision there are
two possible treatment options coded as 0 and 1, that is, $\Psi_1(s_1) =
\mA_1 = \{0,1\}$ for all $s_1$ and $\Psi_2(\bar{s}_2,a_1) =\mA_2 =
\{0,1\}$ for all $\bar{s}_2$ and $a_1\in\{0,1\}$. Let $\mH_1 =
(1,s_1^T)^T$ and $\mH_2 = (1,s^T_1,a_1,s^T_2)^T$. As in many modeling
contexts, it is standard to adopt linear models for the $Q$-functions;
accordingly, consider the models
%
\begin{eqnarray}\label{eq:exampleQ}
Q_1(s_1,a_1; \xi_1) &=&
\mH_1^T\beta_1 + a_1 \bigl(
\mH_1^T \psi_1 \bigr),
\nonumber
\\[-8pt]
\\[-8pt]
\nonumber
 Q_2(
\bar{s}_2,\bar{a}_2; \xi_2) &=&
\mH_2^T\beta_2 + a_2 \bigl(
\mH_2^T \psi_2 \bigr),
\end{eqnarray}
where $\xi_k = (\beta_k^T,\psi^T_k)^T$, $k=1,2$. In
(\ref{eq:exampleQ}), $Q_2(\bar{s}_2,\bar{a}_2;   \xi_2)$ is a model for
$\E(Y | \barS_2=\bar{s}_2,\barA_2=\bar{a}_2)$, a standard regression
problem involving observable data, whereas $Q_1(s_1,a_1; \xi_1)$ is a
model for the conditional expectation of $V_2(\bar{s}_2,a_1) = \max_{a_2
\in\{0,1\}} \E(Y | \barS_2=s_2, A_1=a_1, A_2=a_2)$ given $S_1=s_1$
and $A_1=a_1$, which is an approximation to a complex true
relationship; see  Section~\ref{ss:compare}. Under (\ref{eq:exampleQ}),
$V_2(\bar{s}_2,a_1; \xi_2) =   \max_{a_2 \in\{0,1\}} Q_2(\bar{s}_2,a_1,
a_2; \xi_2) = \mH_2^T\beta_2 + (\mH_2^T \psi_2) \times  I(\mH_2^T \psi_2>0)$
and $V_1(s_1; \xi_1) =\max_{a_1 \in\{0,1\}} Q_1(s_1,a_1;\break
\xi_1) = \mH_1^T\beta_1 + (\mH_1^T \psi_1) I(\mH_1^T \psi_1>0)$.
Substituting the \iQ-functions in (\ref{eq:exampleQ}) in (\ref
{eq:dfinaloptK}) and
(\ref{eq:dfinaloptk}) then yields $d^{\mathrm{opt}}_1(s_1; \xi_1) =
I(\mH_1^T
\psi_1>0)$ and\vspace*{1pt} $d^{\mathrm{opt}}_2(\bar{s}_2,a_1; \xi_2) = I(\mH
_2^T \psi_2
> 0)$.

We have presented (\ref{eq:qlearnK}) and (\ref{eq:qlearnk}) in the
conventional WLS form, with leading term in the summand
$\partial/\partial\xi_k
Q_k(\barS_{ki},\barA_{ki};\xi_k)\Sigma_k^{-1}(\barS_{ki},\barA_{ki})$;
taking $\Sigma_k$ to be a constant yields OLS. At the $K$th decision,
with responses $Y_i$, standard theory implies that this is the optimal
leading term when $\mbox{var}(Y|\barS_K=s_K,\barA_K=a_K) =
\Sigma_K(\bar{s}_K,\bar{a}_K)$, yielding the (asymptotically) efficient
estimator for $\xi_K$. For $k<K$, with ``responses''
$\tilV_{(k+1)i}$, this theory may no longer apply, however, deriving
the optimal leading term involves considerable complication.
Accordingly, it is standard to fit the posited models
$Q_k(\bar{s}_k,\bar{a}_k; \xi_k)$ via OLS or WLS; some authors
define \iQ-learning as using OLS (\citep{Chakraborty}). The choice may
be dictated by apparent relevance of the homoscedasticity
assumption on the $\tilV_{(k+1)i}$, $k=K,K-1,\ldots,1$, and whether
or not
linear models are sufficient to approximate the relationships may also
be evaluated; see Section~\ref{ss:compare}.

\subsection{A-Learning}
\label{ss:alearn}

Advantage learning (\iA-learning, \cite{Blatt}) is a term used to
describe a class of alternative methods to \iQ-learning predicated on
the fact that the entire \iQ-function need not be specified to
estimate the optimal regime. For simplicity, we consider here only
the case of two treatment options coded as 0 and 1 at each decision,
that is, $\Psi_k(\bar{s}_k,\bar{a}_{k-1}) =\mA_k = \{0,1\}$,
$k=1,\ldots,K$.

To fix ideas, consider (\ref{eq:exampleQ}). Note that $d^{\mathrm{opt}}_1(s_1;
\xi_1)$ implied by (\ref{eq:exampleQ}) depends only on $\mH_1^T \psi_1
= Q_1(s_1,1;\break \xi_1)-Q_1(s_1,0; \xi_1)$; likewise,
$d^{\mathrm{opt}}_2(\bar{s}_2,a_1; \xi_2)$ depends only on $\mH
_2^T\psi_2 =
Q_2(\bar{s}_2,a_1,1; \xi_2)- Q_2(\bar{s}_2,a_1,0; \xi_2)$. This reflects
the general result that, for purposes of deducing the optimal regime,
for each $k=1,\ldots,K$, it suffices to know the contrast function
$C_k(\bar{s}_k,\bar{a}_{k-1}) =
Q_k(\bar{s}_k,\bar{a}_{k-1},1)-Q_k(\bar{s}_k,\bar{a}_{k-1},0)$.
This can be
appreciated by noting that any arbitrary $Q_k(\bar{s}_k,\bar{a}_k)$
may be
written as $h_k(\bar{s}_k,\bar{a}_{k-1}) + a_k C_k(\bar{s}_k,\bar
{a}_{k-1})$,
where $h_k(\bar{s}_k,\bar{a}_{k-1})=Q_k(\bar{s}_k,\bar
{a}_{k-1},0)$, so that
$Q_k(\bar{s}_k,\bar{a}_{k-1},  a_k)$ is maximized by taking $a_k = I\{
C_k(\bar{s}_k,\bar{a}_{k-1}) > 0\}$; and the maximum itself is the
expression $h_k(\bar{s}_k,  \bar{a}_{k-1}) + C_k(\bar{s}_k,\bar{a}_{k-1})
I\{C_k(\bar{s}_k,\break \bar{a}_{k-1}) > 0\}$. In the case of two treatment
options we consider here, the contrast function is also referred to as
the optimal-blip-to-zero function (\citep{Robins04}, \citep{Demystifying}).
\citet{Murphy03} considers the expression $C_k(\barS_k,\barA_{k-1})[
I\{ C_k(\barS_k,\barA_{k-1})>0\}-A_k ]$, referred to as the advantage
or regret function, as it represents the ``advantage'' in response
incurred if the optimal treatment at the $k$th decision were given
relative to that actually received (or, equivalently, the ``regret''
incurred by not using the optimal treatment). See \citet{Robins04}
and \citet{Demystifying} for discussion of the relationship between
regrets and optimal blip functions in this and settings other than
binary treatment options.

We discuss here an \iA-learning method based on explicit modeling of
the contrast functions, which we refer to as contrast-based
\iA-learning. This approach is implemented via recursive solution of
certain estimating equations given below developed by \citet{Robins04},
often referred to as g-estimation. See \citet{Demystifying} and the
supplementary material to \citet{Zhang13} for details. Contrast-based
\iA-learning is distinguished from the regret-based \iA-learning
methods of \citet{Murphy03} and \citet{Blatt}, which rely on direct
modeling of the regret functions and are implemented using a different
estimating equation formulation called Iterative Minimization for
Optimal Regimes by \citet{Demystifying}.

All of these methods are alternatives to \iQ-learning, which involves
modeling the full \iQ-func\-tions. For $k=K-1,\ldots,1$, the
\iQ-functions involve possibly complex relationships, raising concern
over the consequences of model misspecification for estimation of the
optimal regime. As identifying the optimal regime depends only on
correct specification of the contrast or regret functions,
\iA-learning methods may be less sensitive to mismodeling; see
Sections~\ref{ss:compare} and \ref{s:simulations}.

Although we consider these methods only in the case of binary
treatment options here, they may be extended to more than two
treatments at the expense of complicating the formulation; see
\citet{Robins04} and \citet{Demystifying}.

Contrast-based \iA-learning proceeds as follows. Posit models
$C_k(\bar{s}_k,\bar{a}_{k-1}; \psi_k)$, $k=1,\ldots,K$, for the contrast
functions, depending on parameters $\psi_k$. Consider decision
$K$. Let $\pi_K(\bar{s}_K, \bar{a}_{K-1}) = \pr(A_K=1|\barS_K=\bar{s}_K,
\barA_{K-1}=\bar{a}_{K-1})$ be the propensity of receiving treatment 1
in the observed data as a function of past history and
$\tilV_{(K+1)i}=Y_i$. \citet{Robins04} showed that all consistent and
asymptotically normal estimators for $\psi_K$ are solutions to
estimating equations of the form
%
\begin{eqnarray}\label{eq:gest}
&&\sum_{i=1}^n
\lambda_K(\barS_{Ki},\barA_{(K-1)i}) \bigl
\{A_{Ki}-\pi_K(\barS_{Ki},\barA_{(K-1)i})
\bigr\}
\nonumber
\\
&&\quad{}\times \bigl\{\tilV_{(K+1)i} - A_{Ki}C_K(
\barS_{Ki},\barA_{(K-1)i};\psi_K) \\
&&\hspace*{94pt}\quad{}-
\theta_K(\barS_{Ki},\barA_{(K-1)i}) \bigr\} = 0\nonumber
\end{eqnarray}
for arbitrary functions $\lambda_K(\bar{s}_K,\bar{a}_{K-1})$ of the same
dimension as $\psi_K$ and arbitrary functions
$\theta_K(\bar{s}_K,\break \bar{a}_{K-1})$. Assuming that the model
$C_K(\bar{s}_K,\bar{a}_{K-1}; \psi_K)$ is correct, if
$\mbox{var}(Y|\barS_K=s_k,\barA_{K-1}=a_{k-1})$ is constant, the
optimal choices of these functions are given by $\lambda_K(\bar
{s}_K,\bar{a}_{K-1};
\psi_K) = \partial/\partial\psi_K C_K(\bar{s}_K,\bar{a}_{K-1};\break
\psi_K)$
and $\theta_K(\bar{s}_{Ki},\bar{a}_{(K-1)i}) = h_K(\bar{s}_K,\bar
{a}_{K-1})$;
otherwise, if the variance is not constant, the optimal $\lambda_K$ is
complex (\citep{Robins04}).

To implement estimation of $\psi_K$ via (\ref{eq:gest}), one may adopt
parametric models for these functions. Although \iA-learning obviates
the need to specify fully the $Q$-functions, one may posit models for
the optimal $\theta_K$, $h_K(\bar{s}_K,\bar{a}_{K-1}; \beta_K)$, say.
Moreover, unless the data are from a SMART study, in which case the
propensities $\pi_K(\bar{s}_K,\bar{a}_{K-1})$ are known, these may be
modeled as $\pi_K(\bar{s}_K, \bar{a}_{K-1}; \phi_K)$ (e.g., by a logistic
regression). These models are only adjuncts to estimating $\psi_K$;
as long as at least one of these models is correctly specified,
(\ref{eq:gest}) will yield a consistent estimator for $\psi_K$, the
so-called double robustness property. In contrast, \iQ-learning
requires correct specification of all \iQ-functions; see
Section~\ref{ss:compare} and Section~A.5 of the supplemental
article [\citet{Schulte}].

Substituting these models in (\ref{eq:gest}), one solves
(\ref{eq:gest}) jointly in $(\psi_K^T,\beta_K^T,\phi_K^T)^T$ with
\begin{eqnarray*}
&&\sum_{i=1}^n \frac{\partial h_K(\barS_K,\barA_{K-1};
\beta_K)}{\partial\beta_K} \\
&&\quad{}\times \bigl\{
\tilV_{(K+1)i} - A_{Ki}C_K(\barS_{Ki},
\barA_{(K-1)i};\psi_K) \\
&&\hspace*{62pt}\qquad{}- h_K(\barS_{Ki},
\barA_{(K-1)i}; \beta_K) \bigr\} = 0
\end{eqnarray*}
and the usual binary regression likelihood score equations
in $\phi_K$. We then have $d^{\mathrm{opt}}_K(\bar{s}_K,\bar
{a}_{K-1};\psi_K) =
I\{ C_K(\bar{s}_K,\bar{a}_{K-1}; \psi_K) > 0 \}$; as in \iQ-learning,
substituting $\hatpsi_K$ yields an estimator for the optimal treatment
choice at decision $K$ for a patient with past history
$\barS_K=s_K,\barA_{K-1}=\bar{a}_{K-1}$.

With $\hatpsi_K$ in hand, the contrast-based
$A$-learning algorithm proceeds in a backward iterative fashion to
yield $\hatpsi_k$, $k=K-1,\ldots,1$. At the $k$th decision, given
models $h_k(\bar{s}_k,\bar{a}_{k-1}; \beta_k)$ and $\pi_k(\bar{s}_k,
\bar{a}_{k-1}; \phi_k)$, one solves jointly in
$(\psi_k^T,\beta_k^T,\phi^T_k)^T$ a system of estimating equations
analogous to those above. The $k$th set of
equations is based on ``optimal responses'' $\tilV_{(k+1)i}$, where,
for each $i$, $\tilV_{ki}$ estimates
$V_k(\barS_{ki},\barA_{(k-1),i})$. It may be shown (see
Section~A.3 of the supplemental article [\citet{Schulte}])
that $E ( V_{k+1}(\barS_{k+1},\barA_k) +
C_k(\barS_k,\barA_{k-1})[ I\{ C_k(\barS_k,\barA_{k-1})>0\}-A_k ]
\rrvert\barS_k,\barA_{k-1} ) =
V_k(\barS_k,\barA_{k-1})$. Accordingly, define recursively
$\tilV_{ki} = \tilV_{(k+1)i} +
C_k(\barS_{ki},\barA_{(k-1)i}; \hatpsi_k)\*[ I\{
C_k(\barS_{ki},\barA_{(k-1)i};   \hatpsi_k)>0\}-A_{ki}]$,
$k=K,K-1,\break \ldots,1$,
$\tilV_{(K+1)i}=Y_i$. The equations at the $k$th decision are then
%
\begin{eqnarray}\label{eq:alearnk}
&&\sum_{i=1}^n
\lambda_k(\barS_{ki},\barA_{(k-1)i};
\psi_k) \bigl\{A_{ki}-\pi_k(
\barS_{ki},\barA_{(k-1)i};\phi_k) \bigr\}
\nonumber
\\
&&\quad{}\times \bigl\{\tilV_{(k+1)i} - A_{ki}C_k(
\barS_{ki},\barA_{(k-1)i};\psi_k)\nonumber\\
&&\hspace*{56pt}\qquad{} - h_k(
\barS_{ki},\barA_{(k-1)i}; \beta_k) \bigr\} = 0,
\nonumber
\\[-6pt]
\\[-6pt]
\nonumber
&&\sum_{i=1}^n \frac{\partial h_k(\barS_K,\barA_{K-1};
\beta_k)}{\partial\beta_k}\\
&&{}\quad\times \bigl\{\tilV_{(k+1)i} - A_{ki}C_k(
\barS_{ki},\barA_{(k-1)i};\psi_k)\nonumber\\
&&\hspace*{56pt}\qquad{} - h_k(
\barS_{ki},\barA_{(k-1)i}; \beta_k) \bigr\} = 0\nonumber
\end{eqnarray}
for a given specification $\lambda_k(\bar{s}_k,\bar{a}_{k-1}; \psi_k)$,
solved  jointly with the maximum likelihood score equations for binary
regression in $\phi_k$. It follows that $d^{\mathrm{opt}}_k(\bar
{s}_k,\bar{a}_{k-1};
\hatpsi_k) = I\{ C_k(\bar{s}_k,\bar{a}_{k-1}; \hatpsi_k) > 0 \}$.
As  abo\-ve, the optimal $\lambda_k$ is complex
(\citep{Robins04}); taking $\lambda_k(\bar{s}_k,\bar{a}_{k-1}; \psi_k)=
\partial/\partial\psi_k C_k(\bar{s}_k,\bar{a}_{k-1}; \psi_k)$ is
reasonable for practical implementation.

Summarizing, the estimated optimal
regime $\widehat{d}^{\mathrm{opt}}_A = (\widehat{d}^{\mathrm
{opt}}_{A,1},\ldots,\widehat{d}^{\mathrm{opt}}_{A,K})$ is
%
\begin{eqnarray}\label{eq:aestdopt}
\widehat{d}^{\mathrm{opt}}_{A,1}(s_1)& =&
d^{\mathrm{opt}}_1(s_1; \hatpsi_1),\nonumber\\
\widehat{d}^{\mathrm{opt}}_{A,k}(\bar{s}_k,
\bar{a}_{k-1}) &= & d^{\mathrm{opt}}_k(\bar{s}_k,a_{k-1};
\hatpsi_k),\\
\eqntext{ k=2,\ldots,K.}
\end{eqnarray}
How well $\widehat{d}^{\mathrm{opt}}_A$ estimates $d^{\mathrm{opt}}$
and hence $d^{(1){\mathrm{opt}}}$ depends
on how close the posited $C_k(\bar{s}_k,\bar{a}_{k-1}; \psi_k)$ are
to the
true contrast functions as well as correct specification of the
functions $h_k$ or $\pi_k$.

Henceforth, for brevity, we suppress the descriptor ``contrast-based''
and refer to the foregoing approach simply as \iA-learning.

\subsection{Comparison and Practical Considerations}
\label{ss:compare}

When $K=1$, the $Q$-function is a model for $\E(Y|  S_1=s_1,A_1=a_1)$.
If in \iQ-learning this model and the variance model $\Sigma_1$ in
(\ref{eq:qlearnK}) are correctly specified, then, as above, the form
of (\ref{eq:qlearnK}) is optimal for estimating~$\xi_1$. Accordingly,
even if $C_1(s_1; \psi_1)$ and $h_1(s_1; \beta_1)$ are correctly
modeled, (\ref{eq:alearnk}) with $K=1$ is generally not of this
optimal form for any choice $\lambda_1(s_1;\psi_1)$, and, hence,
\iA-learning will yield relatively inefficient inference on $\psi_1$
and the optimal regime. However, if in \iQ-learning the $Q$-function
is mismodeled, but in \iA-learning $C_1(s_1; \psi_1)$ and
$\pi_1(s_1;\phi_1)$ are both correctly specified, then \iA-learning
will still yield consistent inference on $\psi_1$ and hence the
optimal regime, whereas inference on $\xi_1$ and the optimal regime
via \iQ-learning may be inconsistent. We assess the trade-off between
consistency and efficiency in this case in
Section~\ref{s:simulations}. For $K>1$, owing to the complications
involved in specifying optimal estimating equations for \iQ- and
\iA-learning, relative performance is not readily apparent; we
investigate empirically in Section~\ref{s:simulations}.

In special cases, \iQ- and \iA-learning lead to identical
estimators for the \iQ-function (Chakraborty, Murphy and Strecher, \citeyear{Chakraborty}). For example, this
holds if the propensities for treatment are constant, as would be the
case under pure randomization at each decision point, and certain
linear models are used for $C_1(s_1; \psi_1)$ and $h_1(s_1; \beta_1)$;
Section~A.4 of the supplemental article [\citet{Schulte}]
demonstrates when $K=1$ and $\pr(A_1=1|S_1=s_1)$ does not
depend on $s_1$. See \citeauthor{Robins04} [(\citeyear{Robins04}), page 1999] and
\citet{Rosenblum} for further discussion.

As we have emphasized, for \iQ-learning, while modeling the
$Q$-function at decision $K$ is a standard regression problem with
response $Y$, for decisions $k=K-1,\ldots,1$, this involves modeling
the estimated value function, which at decision $k$ depends on
relationships for future decisions $k+1,\ldots,K$. Ideally, the
sequence of posited models $Q_k(\bar{s}_k,\bar{a}_k; \xi_k)$ should
respect this constraint. However, this may be difficult to achieve
with standard regression models. To illustrate, consider
(\ref{eq:exampleQ}), and assume $S_1,S_2$ are scalar, where the
conditional distribution of $S_2$ given $S_1=s_1,A_1=a_1$ is
$\mathrm{Normal}(\mK^T_1 \gamma,\sigma^2)$, say, $\mK_1 =
(1,s_1,a_1)^T$. Recall that $V_2(\bar{s}_2,a_1; \xi_2) = \mH^T_2
\beta_2
+ (\mH^T_2 \psi_2) I(\mH^T_2 \psi_2>0)$, where
$\mH^T_2\beta_2 = \mK_1^T\beta_{21} + s_2 \beta_{22}$ and $\mH^T_2
\psi_2 = \mK_1^T\psi_{21} + s_2 \psi_{22}$. Then, if model $Q_2$
in (\ref{eq:exampleQ}) were correct, from (\ref{eq:qk}), ideally,
$Q_1(s_1,a_1) = \E\{V_2(s_1,S_2,a_1; \xi_2)|S_1=s_1,A_1=a_1\}$.
Letting $\varphi(\cdot)$ and $\Phi(\cdot)$ be the standard normal
density and cumulative distribution function, respectively, it may be
shown (see Section~A.5 of the supplemental article
[\citet{Schulte}]) that
%
\begin{eqnarray}\label{eq:realq1}
 Q_1(s_1,a_1) &=& \E \bigl
\{V_2(s_1,S_2,a_1;
\xi_2)|S_1=s_1,A_1=a_1
\bigr\} \nonumber\\
&=& \mK_1^T (\beta_{21} + \gamma
\beta_{22})
\nonumber
\\
&& {}+ \bigl(\mK_1^T\psi_{21} \bigr) \bigl\{ 1-
\Phi(\eta) \bigr\} \\
&&{}+ \psi_{22} \bigl\{\sigma\varphi(\eta) + \bigl(
\mK_1^T\gamma \bigr) \bigl\{ 1-\Phi(\eta) \bigr\} \bigr
\},\nonumber\\
 \eta&=&-\mK_1^T(\psi_{21}/
\psi_{22}+ \gamma)/\sigma, \nonumber
\end{eqnarray}
taking $\psi_{22}>0$. The true $Q_1(s_1,a_1)$ in (\ref{eq:realq1}) is
clearly highly nonlinear and likely poorly approximated by the posited
linear model $Q_1(s_1,a_1;\xi_1)$ in (\ref{eq:exampleQ}). For larger
$K$, this incompatibility between true and assumed models would
propagate from $K-1,\ldots,1$. Thus, while using linear models for
the $Q$-functions is popular in practice, the potential for such
mismodeling should be recognized.

An approach that may mitigate the risk of mismodeling is to employ
flexible models for the \iQ-functions; \citet{Zhao} use support
vector regression models. Developments in statistical learning
suggest a large collection of powerful regression methods that might
be used. Many of these methods must be tuned in order to balance bias
and variance, a natural approach to which is to minimize the
cross-validated mean squared error of the \iQ-functions at each
decision point. An obvious downside is that the final model may be
difficult to interpret, and clinicians may not be willing to
use ``black box'' rules. One compromise is to fit a simple,
interpretable model, such as a decision tree, to the fitted values of
the complex model in order to explore the factors driving the recommended
treatment decisions. This simple model can then be checked against
scientific theory. If it appears sensible, then clinicians may be
willing to use predictions from the complex model. For discussion,
see \citet{Craven}.

\iA-learning represents a middle ground between \iQ-learning and these
approaches in that it allows for flexible modeling of the functions
$h_k(\bar{s}_{k}, \bar{a}_{k-1})$ while maintaining simple parametric
models for the contrast functions $C_k(\bar{s}_{k}, \bar{a}_{k-1})$.
Thus, the resulting decision rule, which depends only on the contrast
function, remains interpretable, while the model for the response is
allowed to be nonlinear. This is also appealing in that it may be
reasonable to expect, based on the underlying science, that the
relationship between patient history and outcome is complex while the
optimal rule for treatment assignment is dependent, in a simple
fashion, on a small number of variables. The flexibility allowed by a
semiparametric model also has its drawbacks. Techniques for formal
model building, critique and diagnosis are well understood for linear
models but much less so for semiparametric models. Consequently,
\iQ-learning based on building a series of linear models may be more
appealing to an analyst interested in formal diagnostics.

\iA-learning may have certain advantages for making inference under
the null hypothesis of no effect of any treatment regime in $\mD$ on
outcome. For example, in a SMART, the propensities are specified by
design, and, under the null, the contrast functions are identically
zero and hence correctly specified. Thus,
\mbox{\iA-learning} will yield
consistent estimators for the parameters defining the contrast
function. See \citet{Robins04} and the references in
Section~\ref{s:discuss}.

\section{Simulation Studies}
\label{s:simulations}

We examine the finite sample performance of \iQ- and \iA-learning on a
suite of simple test examples via Monte Carlo simulation. We
emphasize that the methods are straightforward to implement in more
complex settings than those here. To illustrate trade-offs between
the methods, we begin with correctly specified models and
systematically introduce misspecification of the \iQ-function, the
propensity model and both. We focus here on situations where the
contrast function is correctly specified to gain insight into impact
of other model components. Scenarios with a misspecified contrast
model can be constructed to include or exclude the target $d^{\mathrm{opt}}$,
precluding generalizable conclusions. See Section~A.9 of the
supplemental article [\citet{Schulte}] and
\citeauthor{Stat} (\citeyear{Stat,Zhang12,Zhang13}) for simulations involving misspecified contrast
functions and \citet{Robins04}, Section~9, for discussion.

%
\begin{figure*}[b]

\includegraphics{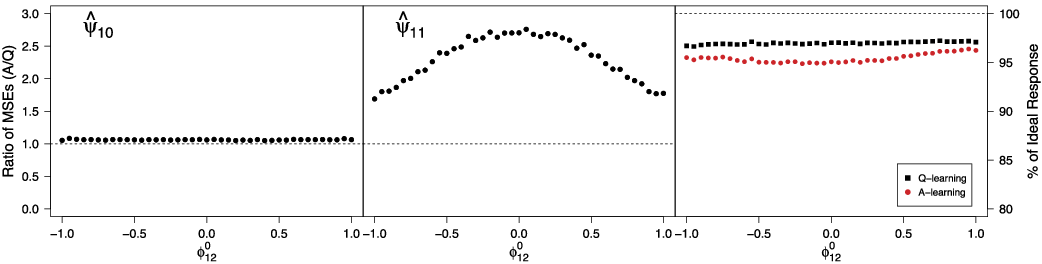}

\caption{Monte Carlo MSE ratios for estimators of components of
$\psi_1$ (left and center panels) and efficiencies $R(\widehat
{d}^{\mathrm{opt}}_Q)$ and
$R(\widehat{d}^{\mathrm{opt}}_A)$ for estimating the true $d^{\mathrm
{opt}}$ (right panel) under
misspecification of the propensity model. MSE ratios $>$ 1 favor \iQ-learning.}
\label{fig1prop}
\end{figure*}

In all scenarios, 10,000 Monte Carlo replications were used, and, for
each generated data set, $\widehat{d}^{\mathrm{opt}}_Q$ and $\widehat
{d}^{\mathrm{opt}}_A$ in
(\ref{eq:qestdopt}) and (\ref{eq:aestdopt}) were obtained using the
\iQ- and \iA-learning procedures in Sections~\ref{ss:qlearn} and~\ref{ss:alearn}.
For simplicity, we consider one ($K=1)$ and two
($K=2$) decision problems, where, at each decision point, there are
two treatment options coded as 0 and 1. In all cases, we used
\iQ-functions of the form $Q_{1}(s_1, a_1;\xi_1) = h_1(s_1;\beta_1) +
a_1 C_1(s_1;\psi_1)$ and $Q_{2}(\bar{s}_2, \bar{a}_{2};\xi_2) =
h_2(\bar{s}_2; a_1;\beta_2) + a_2 C_2(\bar{s}_{2}, a_1; \psi_2)$ to
represent both true and assumed working models. With the contrast
functions correctly specified, $\psi_k$, $k=1,2$, dictate the optimal
regime. Thus, as one measure of performance, we focus on relative
efficiency of the estimators of components of $\psi_k$ as reflected by
the ratio of Monte Carlo mean squared errors (MSEs) given by MSE of
\iA-learning/MSE of \iQ-learning, so that values greater than 1 favor
\iQ-learning. Recognizing that $\E\{Y^*(d^{\mathrm{opt}})\}$ is the benchmark
achievable outcome on average, as a second measure, we consider the
extent to which the estimated regimes $\widehat{d}^{\mathrm{opt}}_Q$
and $\widehat{d}^{\mathrm{opt}}_A$
achieve $\E\{Y^*(d^{\mathrm{opt}})\}$ if followed by the population.
Specifically, for regime $d$ indexed by $\psi_1$ ($K=1$) or
$(\psi_1^T,\psi_2^T)^T$ ($K=2$), let $H(d) = \E\{Y^*(d)\}$, a function
of these parameters. Then $H(d^{\mathrm{opt}})=\E\{Y^*(d^{\mathrm
{opt}})\}$ is this
function evaluated at the true parameter values, and $H(\widehat
{d}^{\mathrm{opt}})$ is
this function evaluated at the estimated parameter values for a given
data set, where $\widehat{d}^{\mathrm{opt}}$ is $\widehat
{d}^{\mathrm{opt}}_Q$ or $\widehat{d}^{\mathrm{opt}}_A$. Define
$R(\widehat{d}^{\mathrm{opt}})=\E\{H(\widehat{d}^{\mathrm{opt}})\}
/H(d^{\mathrm{opt}})$, where the expectation in the
numerator is with respect to the distribution of the estimated
parameters in $\widehat{d}^{\mathrm{opt}}$. We refer to $R(\widehat
{d}^{\mathrm{opt}})$ as the
$v$-efficiency of $\widehat{d}^{\mathrm{opt}}$, as it reflects the extent
to which $\widehat{d}^{\mathrm{opt}}$ achieves the ``value'' of the
true optimal
regime. In Section~A.6 of the supplemental article
[\citet{Schulte}] we discuss calculation of $R(\widehat{d}^{\mathrm{opt}})$.

\subsection{One Decision Point}
\label{ss:onedecision}

In this and the next section, $n=200$. Here, the observed data are
$(S_{1i}, A_{1i}, Y_i)$, $i=1,\ldots,n$. With $\mathrm{expit}(x) =
e^x/(1+e^x)$, we used the class of generative models
%
\begin{eqnarray}\label{eq:generative1}
S_1 &\sim&\mathrm{Normal}(0, 1),\nonumber\\
A_1|S_1&=& s_1 \sim\mathrm{Bernoulli} \bigl
\{\mathrm{expit} \bigl(\phi_{10}^{0} + \phi_{11}^{0}s_1\nonumber\\
&&\hspace*{110pt}{}+ \phi_{12}^{0}s_1^2 \bigr) \bigr
\},
\nonumber
\\[-8pt]
\\[-8pt]
\nonumber
Y|S_1& =&s_1,\\
A_1 &=& a_1 \sim
\mathrm{Normal} \bigl\{\beta_{10}^{0} + \beta_{11}^{0}s_1
+ \beta_{12}^{0}s_1^2\nonumber\\
&&\hspace*{60pt}{} +
a_1 \bigl(\psi_{10}^{0} + \psi_{11}^{0}s_1
\bigr), 9 \bigr\},\nonumber
\end{eqnarray}
indexed by $\phi^0 = (\phi_{10}^{0}, \phi_{11}^{0}, \phi_{12}^{0})^T$,
$\beta^0=(\beta_{10}^{0}, \beta_{11}^{0},   \beta_{12}^{0})^T$,
$\psi^0=(\psi_{10}^{0}, \psi_{11}^{0})^{T}$, so that $d^{\mathrm
{opt}}=d^{\mathrm{opt}}_1$,
$d^{\mathrm{opt}}_1(s_1) = I(\psi^0_{10} + \psi^0_{11} s_1 > 0)$. For
\iA-learning, we assumed models $h_{1}(s_1; \beta_1) = \beta_{10} +
\beta_{11}s_1$, $C_{1}(s_1;\psi_1) = \psi_{10} + \psi_{11}s_1$, and
$\pi_{1}(s_1;\phi_1) = \expit(\phi_{10} + \phi_{11}s_1)$, and for
\iQ-learning we used $Q_1(s_1, a_1; \xi_1) = h_{1}(s_1; \beta_1) + a_1
C_{1}(s_1;\psi_1)$. These models involve correctly specified contrast
functions and are nested within the true models, with
$h_{1}(s_1;\beta_1)$, and hence the \iQ-function, correctly specified
when $\beta_{12}^{0} = 0$. The propensity model $\pi_{1}(s_1;\phi_1)$
is correctly specified when $\phi_{12}^{0} = 0$. To study the effects
of misspecification, we varied $\beta_{12}^{0}$ and $\phi_{12}^{0} $
while keeping the others fixed, considering parameter settings of the
form $\phi^0=(0, -2,\phi_{12}^{0})^T$,
$\beta^0=(1,1,\beta_{12}^{0})^T$, $\psi^0=(1, 0.5)^{T}$.

\textit{Correctly specified models}.
\label{sss:oneCorrect}
As noted in Section~\ref{ss:compare}, when all working models are
correctly specified, \iQ-learning is more efficient than \iA-learning,
which for (\ref{eq:generative1}) occurs when $\beta_{12}^{0} =
\phi_{12}^{0} = 0$. Here, the efficiency of $Q$-learning relative to
$A$-learning is $1.06$ for estimating $\psi_{10}^{0}$ and $2.74$ for
$\psi_{11}^{0}$. Thus, $Q$-learning is a modest 6\% more efficient in
estimating $\psi_{10}^{0}$ but a dramatic 174\% more efficient in
estimating $\psi_{11}^{0}$. Interestingly, the $v$-efficiency of the
decision rules produced by the methods is similar, with $R(\widehat
{d}^{\mathrm{opt}}_Q)
= 0.97$ and $R(\widehat{d}^{\mathrm{opt}}_A) = 0.95$, so that
inefficiency in
estimation of $\psi_1$ via \iA-learning does not translate into a
regime of poorer quality than that found by \iQ-learning.

%
\begin{figure*}[t]

\includegraphics{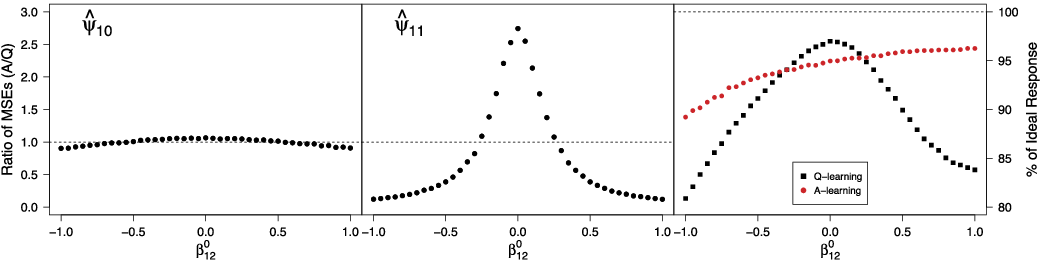}

\caption{Monte Carlo MSE ratios for estimators of components of
$\psi_1$ (left and center panels) and efficiencies $R(\widehat
{d}^{\mathrm{opt}}_Q)$ and
$R(\widehat{d}^{\mathrm{opt}}_A)$ for estimating the true $d^{\mathrm
{opt}}$ (right panel) under
misspecification of the \iQ-function. MSE ratios $>$ 1 favor
\iQ-learning.}
\label{fig1reg}
\end{figure*}

%
\begin{figure*}[b]

\includegraphics{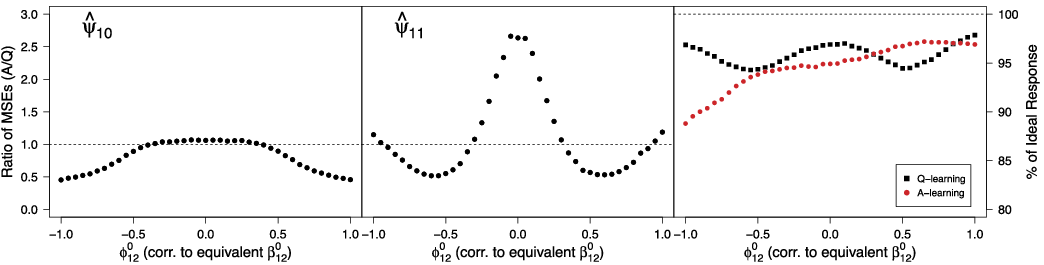}

\caption{Monte Carlo MSE ratios for estimators of components of
$\psi_1$ (left and center panels) and efficiencies $R(\widehat
{d}^{\mathrm{opt}}_Q)$ and
$R(\widehat{d}^{\mathrm{opt}}_A)$ for estimating the true $d^{\mathrm
{opt}}$ (right panel) under
misspecification of both the propensity model and the \iQ-function.
MSE ratios $>$ 1 favor \iQ-learning.}
\label{fig1both}
\end{figure*}

\textit{Misspecified propensity model.} Under (\ref{eq:generative1}),
this situation corresponds to $\beta_{12}^{0} = 0$ and nonzero
$\phi_{12}^{0}$. An appeal of \iA-learning is the double robustness
property noted in Section~\ref{ss:alearn}, which implies that $\psi_1$
is estimated consistently when the propensity model is misspecified
provided that the \iQ-function is correct. In contrast, \iQ-learning
does not depend on the propensity model, so its performance is
unaffected. Figure~\ref{fig1prop} shows the relative efficiency in
estimating $\psi_{10}^{0}$ and $\psi_{11}^{0}$ and the efficiency of
$\widehat{d}^{\mathrm{opt}}_Q$ and $\widehat{d}^{\mathrm{opt}}_A$
as $\phi_{12}^{0}$ varies from $-1$ to
$1$. The leftmost panel shows that there is minimal efficiency gain
by using \iQ-learning instead of \iA-learning in estimation of
$\psi_{10}^{0}$. From the center panel, \iQ-learning yields
substantial gains over $A$-learning for estimating $\psi_{11}^{0}$.
Interestingly, the gain is largest when $\phi_{12}^{0} = 0$, which
corresponds to a correctly specified propensity model. Letting
$\pi^0(s_1; \phi^0_1)$ be the true propensity,
$\phi^0_1=(\phi^0_{10},\phi^0_{11},\phi^0_{12})^T$, a possible
explanation for this seemingly contradictory result in this scenario
is that, as $|\phi_{12}^{0}|$ gets larger, $\logit\{\pi^0(S_1;
\phi^{0}_1)\}=\phi_{10}^{0} + \phi_{11}^{0}s_1 + \phi_{12}^{0}s_1^2$
becomes more profoundly quadratic. Consequently, the estimator for
$\phi_{11}$ in the posited model $\pi_{1}(s_1;\phi_1) =
\expit(\phi_{10} + \phi_{11}s_1)$ approaches zero, so that the
estimated posited propensity approaches a constant. Because \iQ- and
\iA-learning are algebraically equivalent under constant propensity
here, substituting an estimated propensity that is nearly constant
leads to an estimator very similar to that from \iQ-learning.
Consequently, empirical efficiency gains decrease as $|\phi_{12}^{0}|
\rightarrow\infty$. The right panel of Figure~\ref{fig1prop} shows a
small gain in $v$-efficiency of $\widehat{d}^{\mathrm{opt}}_Q$ over
$\widehat{d}^{\mathrm{opt}}_A$; both
achieve good performance.

See Section~A.9 of the supplemental article [\citet{Schulte}]
for evidence demonstrating this behavior of the propensity score and
for further summaries reflecting the relative efficiency of the
estimated regimes in all scenarios in this and the next section.

\textit{Misspecified \iQ-function.} This scenario examines the second
aspect of \iA-learning's double-robustness, characterized in
(\ref{eq:generative1}) by $\phi_{12}^{0} = 0$ and nonzero
$\beta_{12}^{0}$. Here, \iA-learning leads to consistent estimation
while \iQ-learning need not. The left panel of Figure~\ref{fig1reg}
shows that the gain in efficiency using \iA-learning is minimal in
estimating $\psi_{10}^{0}$. The center panel illustrates the
bias-variance trade-off associated with \iQ- versus \iA-learning. For
$\beta_{12}^{0}$ far from zero, bias in the misspecified \iQ-function
dominates the variance, and \iA-learning enjoys smaller MSE while, for
small values of $\beta_{12}^{0}$, variance dominates bias, and
\iQ-learning is more efficient. The right panel shows that large bias
in the \iQ-function can lead to meaningful loss ($\sim$10\%) in
$v$-efficiency of $\widehat{d}^{\mathrm{opt}}_Q$ relative to $\widehat
{d}^{\mathrm{opt}}_A$.

\textit{Both propensity model and $Q$-function misspecified.} In our
class of generative models (\ref{eq:generative1}), this corresponds to
nonzero values of both $\beta_{12}^{0}$ and $\phi_{12}^{0}$. Rather
than vary both values, (e.g., over a grid), we varied one and chose
the other so that it is ``equivalently misspecified.'' In particular,
for a given value of $\phi_{12}^{0}$, we selected $\beta_{12}^{0} =
\beta_{12}^{0}(\phi_{12}^{0})$ so that the $t$-statistic associated
with testing $\phi_{12}^{0} = 0$ in the logistic propensity model and
the $t$-statistic associated with testing $\beta_{12}^{0} = 0$ in the
linear \iQ-function would be approximately equal in distribution.
Consequently, across data sets, an analyst would be equally likely to
detect either form of misspecification. Details of this construction
are given in Section~A.7 of the supplemental article
[\citet{Schulte}].

As in the preceding scenario, Figure~\ref{fig1both} illustrates the
bias-variance trade-off associated with \iQ- and \iA-learning. For
large misspecification, \iA-learning provides a large enough reduction
in bias to yield lower MSE; for small misspecification, \iQ-learning
incurs some bias but reduces the variance enough to yield lower MSE.
From the right panel of the figure, bias seems to translate into a
larger loss in $v$-efficiency of the estimators of $d^{\mathrm{opt}}$
than variance.


\subsection{Two Decision Points}
\label{ss:twodecision}

For $K=2$, the observed data available to estimate $d^{\mathrm{opt}}=
(d^{\mathrm{opt}}_1,d^{\mathrm{opt}}_2)$ are $(S_{1i}, A_{1i},
S_{2i}, A_{2i}, Y_i)$, $i =
1,\ldots,n$. For these scenarios, we used a class of true generative
data models that differs from those of \citet{Chakraborty},
\citet{Song} and \citet{Laberetal} only in that $S_2$ is continuous
instead of binary; as the model at the first stage is saturated, this
allows correct specification of the \iQ-function at decision 1. The
generative model is
\begin{eqnarray*}
&& S_1\sim \mathrm{Bernoulli}(0.5),\\
&&A_1|S_1
= s_1 \sim\mathrm{Bernoulli} \bigl\{\expit \bigl(
\phi_{10}^{0} + \phi_{11}^{0}s_1
\bigr) \bigr\},
\\
&& S_2 | S_1 = s_1,\\
&&A_1 =
a_1 \sim\mathrm{Normal} \bigl( \delta_{10}^{0}
+ \delta_{11}^{0}s_1 + \delta_{12}^{0}a_1
+ \delta_{13}^{0}s_1a_1, 2 \bigr),
\\
 && A_2 | S_1 = s_1,\quad
 S_2= s_2,\\
&& A_1= a_1 \sim\mathrm{Bernoulli} \bigl\{\expit \bigl(
\phi_{20}^{0}+\phi_{21}^{0}s_1
+ \phi_{22}^{0}a_1 \\
&&\hspace*{103pt}{}+ \phi_{23}^{0}s_2
+ \phi_{24}^{0}a_1s_2 +
\phi_{25}^{0}s_2^2 \bigr) \bigr\},
\\
&& Y| S_1= s_1, \quad S_2 = s_2,\\
&&A_1= a_1,\\
&&  A_2=a_2 \sim
\mathrm{Normal} \bigl\{m(s_1,s_2, a_1,a_2),10
\bigr\},
\\
&& m(s_1,s_2,a_1,a_2) =
\beta_{20}^{0} + \beta_{21}^{0}s_1
+ \beta_{22}^{0}a_1\\
&&\hspace*{72pt}\quad{} + \beta_{23}^{0}s_1a_1
+ \beta_{24}^{0}s_2 + \beta_{25}^{0}s_2^2
\\
&&\hspace*{72pt}\quad{}+ a_2 \bigl(\psi_{20}^{0} +
\psi_{21}^{0}a_1 + \psi_{22}^{0}s_2
\bigr).
\end{eqnarray*}
The model is indexed by $\phi_1^{0} =
(\phi_{10}^{0}, \phi_{11}^{0})^{T}$,
$\delta^0_1=(\delta^0_{10},\delta^0_{11},\delta^0_{12},\delta^0_{13})^T$,\vspace*{1pt}
$\phi^0_2=(\phi^0_{20},\phi^0_{21},\phi^0_{22},\phi^0_{23},
\phi^0_{24},  \phi^0_{25})^T$,
$\beta^0_2=(\beta^0_{20},\beta^0_{21},\beta^0_{22},
\beta^0_{23},\beta^0_{24},\beta^0_{25})^T$, and\vspace*{1pt}
$\psi^0_2=(\psi^0_{20}, \psi^0_{21},\break \psi^0_{22})^T$, with true\vspace*{1pt}
$h_{2}^{0}(s_1, s_2, a_1) = \beta_{20}^{0} + \beta_{21}^{0}s_1 +
\beta_{22}^{0}a_1 + \beta_{23}^{0}s_1a_1 + \beta_{24}^{0}s_2 +
\beta_{25}^{0}s_2^2$ and contrast function $C_{2}^{0}(s_1,\allowbreak s_2,\allowbreak  a_1) =
\psi_{20}^{0} + \psi_{21}^{0}a_1 + \psi_{22}^{0}s_2$, say. Because
$A_1$ and $S_1$ are binary, the true functions $h_{1}^{0}(s_1) =
\beta_{10}^{0} + \beta_{11}^{0}s_1$ and $C_{1}^{0}(s_1) =
\psi_{10}^{0} + \psi_{11}^{0}s_1$ are linear in $s_1$;
$\beta_{10}^{0}, \beta_{11}^{0}, \psi_{10}^{0}$ and $\psi_{11}^{0}$
are derived in terms of parameters indexing the generative model in
Section~A.8 of the supplemental article [\citet{Schulte}]. Thus, the
true optimal regime has
$d^{\mathrm{opt}}_1(s_1) = I(\psi_{10}^{0} + \psi_{11}^{0}s_1 > 0)$ and
$d^{\mathrm{opt}}_2(s_1, s_2, a_1) = I(\psi_{20}^{0} + \psi
_{21}^{0}a_1 +
\psi_{22}^{0}s_2 > 0)$.

%
\begin{figure*}[t]

\includegraphics{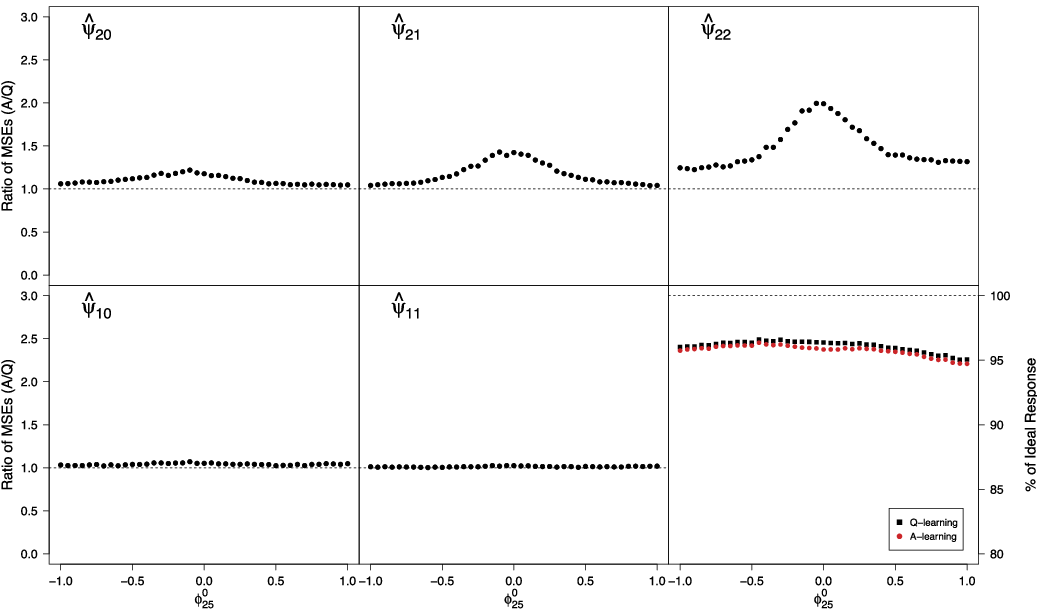}

\caption{Monte Carlo MSE ratios for estimators of components of
$\psi_2$ and $\psi_1$ (upper row and lower row left and center
panels) and
efficiencies $R(\widehat{d}^{\mathrm{opt}}_Q)$ and $R(\widehat
{d}^{\mathrm{opt}}_A)$ for estimating the
true $d^{\mathrm{opt}}$ (lower right panel) under misspecification of
the propensity
model. MSE ratios $>$ 1 favor \iQ-learning.}
\label{fig2prop}
\end{figure*}

We assumed working models for \iA-learning of the form
$h_{1}(s_1;\beta_1) = \beta_{10} + \beta_{11}s_1$,
$C_{1}(s_1;\psi_{1}) = \psi_{10} + \psi_{11}s_1$, $\pi_{1}(s_1;\phi_1)
= \expit(\phi_{10} + \phi_{11}s_1)$, $h_{2}(s_1, s_2, a_1;\break  \beta_2) =
\beta_{20} + \beta_{21}s_1 + \beta_{22}a_1 + \beta_{23}s_1a_1 +
\beta_{24}s_2$, $C_2(s_1,s_2,\break  a_1; \psi_2) =
\psi_{20}+\psi_{21}a_1+\psi_{22}s_2$, and $\pi_{2}(s_1, s_2,
a_1; \phi_2) = \expit(\phi_{20} + \phi_{21}s_1 + \phi_{22}a_1 +
\phi_{23}s_2 + \phi_{24}a_1s_2)$; and, similarly,
\iQ-functions $Q_{1}(s_1,\allowbreak a_1;\allowbreak \xi_1) = h_{1}(s_1;\allowbreak \beta_1) +
a_1C_{1}(s_1;\allowbreak\psi_{1})$ and $Q_{2}(s_1,\allowbreak s_2,\allowbreak a_1,\allowbreak a_2;\allowbreak \xi_2) = h_{2}(s_1,\allowbreak
s_2,\allowbreak  a_1;\allowbreak \beta_2) +a_2 C_2(s_1,\allowbreak s_2,\allowbreak a_1;\allowbreak  \psi_2)$ for \iQ-learning, so
that the contrast functions are correctly specified in each case.
Comparison of the working and generative models shows that the former
are correctly specified when $\phi_{25}^{0}$ and $\beta_{25}^{0}$ are
both zero and are misspecified otherwise. Thus, we systematically varied
these parameters to study the effects of misspecification, leaving all
other parameter values fixed, taking $\phi^0_1=(0.3,-0.5)^T$,
$\delta^0_1=(0,0.5,-0.75,0.25)^T$,
$\phi^0_2=(0,0.5,0.1,-1,-0.1,\phi^0_{25})^T$,\vspace*{1pt}
$\beta^0_2=(3,0,0.1,-0.5,  -0.5,\break \beta^0_{25})^T$, and
$\psi^0_2=(1,0.25,0.5)^T$.

\textit{Correctly specified models.} This occurs when $\phi_{25}^{0}
= \beta_{25}^{0} = 0$. As discussed previously, \iQ-learning is
efficient when the models are correctly specified. Efficiencies of
\iQ- learning relative to \iA-learning for estimating $\psi_{10}^{0}$,
$\psi_{11}^{0}$, $\psi_{20}^{0}$, $\psi_{21}^{0}$ and $\psi_{22}^{0}$
are $1.07$, $1.03$, $1.19$, $1.44$ and $1.98$, respectively. Hence,
\iQ-learning is markedly more efficient in estimating the second stage
parameters but only modestly so for first stage parameters. More
efficient estimators of the parameters do not translate into greater
$v$-efficiency of the estimated regimes in this scenario, as
$R(\widehat{d}^{\mathrm{opt}}_Q)=0.96$ and $R(\widehat{d}^{\mathrm
{opt}}_A)=0.96$.

\textit{Misspecified propensity model.} The propensity  model at the
second stage is misspecified when $\phi_{25}^{0}$ is nonzero. To
isolate the effects of such misspecification, we set $\beta_{25}^{0} =
0$ and varied $\phi_{25}^{0}$ between $-1$ and $1$. From
Figure~\ref{fig2prop}, \iQ-learning is more efficient than
\iA-learning for estimation of all parameters in $\psi_1$ and
$\psi_2$, and, as in the one decision case, the efficiency gain is
largest when $\phi_{25}^{0} = 0$, corresponding to a correctly
specified propensity model. From the lower right panel, there appears
to be little difference in $v$-efficiency of $\widehat{d}^{\mathrm
{opt}}_Q$ and
$\widehat{d}^{\mathrm{opt}}_A$.

\begin{figure*}[t]

\includegraphics{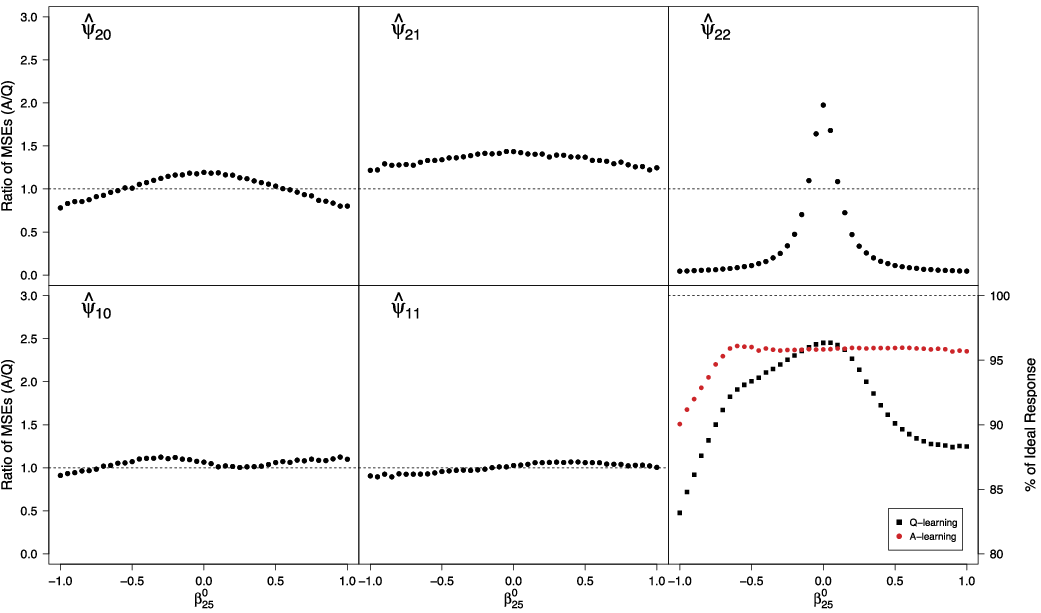}

\caption{Monte Carlo MSE ratios for estimators of components of
$\psi_2$ and $\psi_1$ (upper row and lower row left and center
panels) and
efficiencies $R(\widehat{d}^{\mathrm{opt}}_Q)$ and $R(\widehat
{d}^{\mathrm{opt}}_A)$ for estimating the
true $d^{\mathrm{opt}}$ (lower right panel) under misspecification of the
\iQ-functions. MSE ratios $>$ 1 favor \iQ-learning.}\vspace*{5pt}
\label{fig2reg}
\end{figure*}

\textit{Misspecified $Q$-function.} Under our class of generative
models, the \iQ-function is misspecified when $\beta_{25}^{0}$ is
nonzero. We set $\phi_{25}^{0} = 0$ to focus on the effects of such
misspecification. Figure~\ref{fig2reg} shows that, for the first
stage parameters $\psi_{10}^{0}$ and $\psi_{11}^{0}$, there is little
difference in efficiency between \iQ- and \iA-learning. The upper
panels illustrate varying degrees of the bias-variance trade-off
between the methods. In particular, in estimating $\psi_{22}^{0}$, a
small amount of misspecification leads to significant bias, and, hence,
\iA-learning produces a much more accurate estimator, while, for
$\psi_{20}^{0}$, the bias-variance trade-off is present but
attenuated and there is little difference between \iQ- and
\iA-learning. In estimation of $\psi_{21}^{0}$, variance appears
to dominate bias, and \iQ-learning is preferred for the chosen range
of $\beta_{25}^{0}$ values. From the lower right panel, relative
efficiency for estimating $\psi_{22}^{0}$ weakly tracks the relative
efficiencies of the estimated regimes $\widehat{d}^{\mathrm{opt}}_Q$
and $\widehat{d}^{\mathrm{opt}}_A$,
suggesting that the efficiency gain for \iA-learning in estimating
$\psi_{22}^{0}$ leads to improved estimation of $d^{\mathrm{opt}}$.

%

\textit{Both the propensity model and $Q$-function misspecified.}
This scenario corresponds to nonzero values of $\beta_{25}^{0}$ and
$\phi_{25}^{0}$. Analogous to the one decision case, we chose pairs
$(\beta_{25}^{0}, \phi_{25}^{0})$ that are ``equivalently
misspecified;'' see Section~A.7 of the supplemental article
[\citet{Schulte}]. From Figure~\ref{fig2both}, there is no general
trend in efficiency of estimation across parameters that might
recommend one method over the other. Furthermore, from the lower
right panel, there is little difference in $v$-efficiency of the estimated
regimes. One should not expect to draw broad conclusions, as neither
\iQ- nor \iA-learning need be consistent here. Interestingly, despite
misspecification of both models, $\widehat{d}^{\mathrm{opt}}_Q$ and
$\widehat{d}^{\mathrm{opt}}_A$ still
enjoy high $v$-efficiency in this scenario.

%
\begin{figure*}[t]

\includegraphics{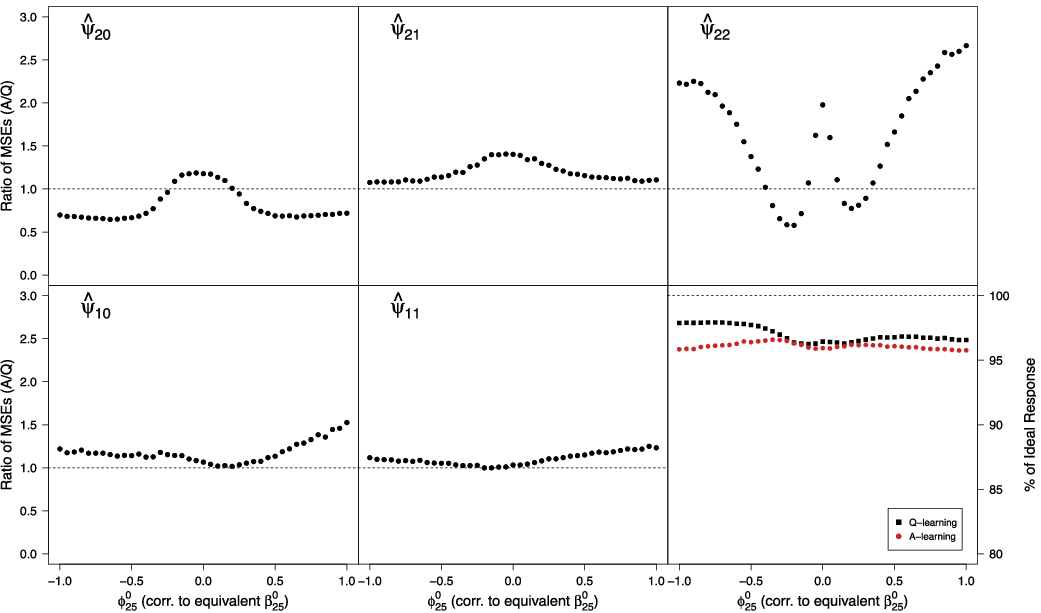}

\caption{Monte Carlo MSE ratios for estimators of components of
$\psi_2$ and $\psi_1$ (upper row and lower row left and center panels)
and efficiencies $R(\widehat{d}^{\mathrm{opt}}_Q)$ and $R(\widehat
{d}^{\mathrm{opt}}_A)$\vspace*{1pt} for estimating
the true $d^{\mathrm{opt}}$ (lower right panel) under misspecification
of both
the propensity models and \iQ-functions. MSE ratios $>$ 1 favor
\iQ-learning.}
\label{fig2both}
\end{figure*}


\subsection{Moodie, Richardson and Stephens Scenario}
\label{ss:moodie}

The foregoing simulation scenarios deliberately involve simple models
for the \iQ-functions in order to allow straightforward interpretation.
To investigate the relative performance of the methods in a more
challenging setting, we generated data from a scenario similar to that
in \citet{Demystifying} in which the true contrast functions are simple
yet the \iQ-functions are complex.



The data generating process used mimics a study in which HIV-infected
patients are randomized~to receive antiretroviral therapy (coded as 1)
or not (coded as 0) at baseline and again at six months, where the
randomization probabilities depend on baseline and six month CD4
counts. Specifically, we generated baseline CD4 count $S_1 \sim
\mathrm{Normal}(450,\break 150^2)$, and baseline treatment $A_1$ was then
assigned according to $A_1 |S_1=s_1 \sim
\mathrm{Bernoulli}\{\expit(\phi^0_{10}+\phi^0_{11}s_1)\}$. We\vspace*{1pt}
generated six month CD4 count $S_2$,\break distributed conditional on
$S_1=s_1, A_1=a_1$ as\break Normal$(1.25 s_1,60^2)$. Treatment $A_2$ was
then generated according to $A_2 | S_1=s_1,A_1=a_1,S_2=s_2 \sim
\mathrm{Bernoulli}\{\expit(\phi^0_{20}+\phi^0_{21}s_2)\}$. In
contrast to the scenario in \citet{Demystifying}, this allows all
possible treatment combinations. The outcome $Y$ is CD4 count at one
year; following \citet{Demystifying}, $Y$ was generated as
$Y=Y^{\mathrm{opt}}-\mu^0_1(S_1,A_1)-\mu^0_2(S_1,S_2,\break  A_1,A_2)$, where
$Y^{\mathrm{opt}}|S_1=s_1,A_1=a_1,S_2=s_2,A_2=a_2 \sim\mathrm{Normal}(400+1.6
s_1,60^2)$. Here, $\mu^0_1(S_1,A_1)$ and   $\mu^0_2(S_1,S_2,A_1,A_2)$ are
the true advantage (regret) functions; we took
$C^0_1(s_1)=\psi^0_{10}+\psi^0_{11}s_1$ and
$C^0_2(s_1,s_2,  a_1)=\psi^0_{20}+\psi^0_{21}s_2$ to be the true contrast
functions, so that, from Section~\ref{ss:alearn},
%
\begin{eqnarray}\quad
\label{eq:moodiereg1}&&\mu^0_1(S_1,A_1)
\nonumber
\\[-8pt]
\\[-8pt]
\nonumber
&&\quad=  \bigl(
\psi^0_{10}+\psi^0_{11}S_1
\bigr) \bigl\{I \bigl(\psi^0_{10}+\psi^0_{11}S_1
>0 \bigr) - A_1 \bigr\},
\\
\label{eq:moodiereg2}&&\mu^0_2(S_1,S_2,A_1,A_2)
\nonumber
\\[-8pt]
\\[-8pt]
\nonumber
&&\quad =  \bigl(\psi^0_{20}+\psi^0_{21}S_2
\bigr) \bigl\{I \bigl(\psi^0_{20}+\psi^0_{21}S_2
>0 \bigr) - A_2 \bigr\}.
\end{eqnarray}
It follows that the optimal treatment regime $d^{\mathrm
{opt}}=(d^{\mathrm{opt}}_1,d^{\mathrm{opt}}_2)$
has $d^{\mathrm{opt}}_1(s_1) = I(\psi^0_{10}+\psi^0_{11}s_1 >0)$ and
$d^{\mathrm{opt}}_2(\bar{s}_2,a_1) = I(\psi^0_{20}+\psi^0_{21}s_2
>0)$. While the
true contrast functions are linear in $\psi^0_k$, $k=1,2$, the true
implied $h^0_1(s_1)$ and $h^0_2(s_1,a_1,s_2)$ are nonsmooth and
possibly complex.

Following\hspace*{-0.5pt} Moodie,\hspace*{-0.5pt} Richardson\hspace*{-0.5pt} and\hspace*{-0.5pt} Stephens\hspace*{-0.5pt} (\citeyear{Demystifying}), for \iA-learning, we assumed working
models $h_1(s_1;\allowbreak \beta_1) = \beta_{10} + \beta_{11} s_1$, $C_1(s_1;
\psi_1) = \psi_{10} + \psi_{11} s_1$, $h_2(s_1,  s_2,\break a_1; \beta_2)
=\beta_{20} + \beta_{21} s_1 + \beta_{22} a_1 +\beta_{23} s_1 a_1 +
\beta_{24} s_2$, and $C_2(s_1,s_2, a_1;\psi_2) =\psi_{20} + \psi_{21}
s_2$, and propensity models $\pi_1(s_1; \phi_1) = \expit(\phi_{10} +
\phi_{11} s_1)$ and $\pi_2(s_1,s_2,   a_1; \phi_2)
=\expit(\phi_{20}+\phi_{21}s_2)$. For \iQ-learning, we analogously assumed
\iQ-functions $Q_1(s_1,a_1; \xi_1) = h_1(s_1;   \beta_1) + a_1 C_1(s_1;
\psi_1)$ and $Q_2(s_1,s_2,a_1,a_2; \xi_2) =h_2(s_1,  s_1,a_1; \beta_2) +
a_2 C_2(s_1,s_2,a_1;\psi_2)$. Note that the contrast functions in
each case are correctly specified, as are the propensity models;
however, the \iQ-functions are misspecified, as the linear models
$h_1(s_1; \beta_1)$ and $h_2(s_1,s_1,a_1; \beta_2)$ are poor
approximations to the complex forms of the true $h^0_1(s_1)$ and
$h^0_2(s_1,s_2,a_1)$.

We report results for $n=1000$ with
$\phi^0_1=(\phi^0_{10},\break  \phi^0_{11})^T=(2.0,-0.006)^T$,
$\phi^0_2=(\phi^0_{20},\phi^0_{21})^T=\vspace*{1pt}(0.8,\break  -0.004)^T$,
$\psi^0_1=(\psi^0_{10},\psi^0_{11})^T=(250,-1.0)^T$, and
$\psi^0_2=(\psi^0_{20},\psi^0_{21})^T=(720,-2.0)^T$ in
Table~\ref{t:moodie}.
%
\begin{table}[b]
\tabcolsep=0pt
\caption{Monte Carlo average (standard deviation) of estimates
obtained via \iQ- and \iA-learning and ratio of Monte Carlo MSE for
the Moodie and Richardson scenario;\break MSE ratios $>$ 1 favor
\iQ-learning}
\label{t:moodie}
\begin{tabular*}{\columnwidth}{@{\extracolsep{\fill}}lccc@{}}
\hline
\textbf{Parameter} &  &
&  \\
\textbf{(true value)} & \textbf{\iQ-learning} &
\textbf{\iA-learning} & \textbf{MSE ratio} \\
\hline
$\psi_{10}^{0}=250$ & 154.8 (23.2) & 249.1 (18.7) & 0.036 \\[2pt]
$\psi_{11}^{0}= -1.0$ & $-$0.775 (0.052) & $-$0.998 (0.041) & 0.032 \\[2pt]
$\psi_{20}^{0}=720$ & 507.3 (49.2) & 720.3 (48.4) & 0.050 \\[2pt]
$\psi_{21}^{0}= -2.0$ & $-$1.584 (0.092) & $-$2.001 (0.085) & 0.040 \\
\hline
\end{tabular*}
\end{table}
Because the \iQ-functions are misspecified, the
\iQ-learning estimators for $\psi^0_1$ and $\psi^0_2$ are biased,
while those obtained via \iA-learning are consistent owing to the
double robustness property. This leads to the dramatic relative
inefficiency of \iQ-learning reflected by the MSE ratios. Under the
assumed models, the estimated optimal regime for \iQ-learning dictates
that, at baseline, therapy be given to patients with
baseline CD4 count less than 199.7, while that estimated using
\iA-learning gives treatment to those with baseline CD4 count less
than 249.1, almost perfectly achieving the true optimal CD4 threshold
of 250. Under the data generative process, using the baseline
decision rule estimated via \iQ-learning may result in as many as
4.4\% of patients who would receive therapy at baseline under the true
optimal regime being assigned no treatment. Similarly, at the second
decision, the estimated optimal regimes obtained by \iQ- and
\iA-learning dictate that therapy be given to patients with six month
CD4 count less than 320.2 and 360.1, respectively. Again,
\iA-learning yields an estimated threshold almost identical to the
optimal value of 360. Although that obtained via \iQ-learning is
lower, 4.3\% of patients who should receive therapy at six months
would not if the estimated six month rule from \iQ-learning were
followed by the population.

By Section~A.6 of the supplemental article [\citet{Schulte}],
$H(d^{\mathrm{opt}}) = 1120$, whereas $\E\{H(\widehat{d}^{\mathrm
{opt}}_Q)\} \approx1117.1$
(estimated standard error 1.3) and\break  $\E\{H(\widehat{d}^{\mathrm
{opt}}_A)\} \approx
1119.9$ (0.3), so that $R(\widehat{d}^{\mathrm{opt}}_Q)$ and
$R(\widehat{d}^{\mathrm{opt}}_A)$ are
virtually equal to one. Thus, although \iQ-learning yields poor
estimation of parameters in the contrast functions, loss in
$v$-efficiency of the estimated optimal regime is negligible. A
possible explanation is as follows. For (\ref{eq:moodiereg1}) and
(\ref{eq:moodiereg2}), some patients near the true treatment decision
boundary would have $C^0_k(\barS_k,\barA_{k-1})$, $k=1,2$, close to
zero. Thus, even if a regime improperly assigns treatment to these
patients, they would experience only a small loss in outcome and hence
have little effect on the overall average. For other patients for
whom the true contrast is not close to zero, improper assignment could
result in considerable degradation of outcome. Because the proportion
of patients receiving improper assignment is small in this
scenario, the effect of these latter patients on the overall expected
outcome is not substantial, leading to the relatively good expected
outcome under the estimated \iQ-learning regime.


\section{Application to STAR*D}
\label{s:application}

Sequenced Treatment Alternatives to Relieve Depression (STAR*D) was a
randomized clinical trial enrolling 4041 patients designed to compare
treatment options for patients with major depressive disorder.
The trial involved four levels, where each level consisted of a 12
week period of treatment, with scheduled clinic visits at weeks 0, 2,
4, 6, 9, 12. Severity of depression at any visit was assessed using
clinician-rated and self-reported versions of the Quick Inventory of
Depressive Symptomatology (QIDS) score (\citep{Rush03}), for which
higher values correspond to higher severity. At the end of each
level, patients deemed to have an adequate clinical response to that
level's treatment did not move on to future levels, where adequate
response was defined by 12-week clinician-rated QIDS score $\le5$
(remission) or showing a 50\% or greater decrease from the baseline
score at the beginning of level 1 (successful reduction). During
level 1, all patients were treated with citalopram. Patients
continuing to level 2 due to inadequate response, conferring with
their physicians, expressed preference to (i) switch or (ii) augment
citalopram and within that preference were randomized to one of
several options: (i) switch: sertraline, bupropion, venlafaxine, or
cognitive therapy, or (ii) augment: citalopram plus one of either
bupropion, buspirone, or cognitive therapy. Patients randomized to
cognitive \mbox{therapy} (alone or augmented with citalopram) were eligible,
in the case of inadequate response, to move to a supplementary level
2A and be randomized to switch to bupropion or venlafaxine. All
patients without adequate response at level 2 (or 2A)
continued to level~3 and, depending on preference to (i) switch or
(ii) augment, were randomized within that preference to (i) switch:
mirtazepine or nortriptyline or (ii) augment with either: lithium or
triiodothyronine. Patients without adequate response continued to
level 4, requiring a switch to tranylcypromine or mirtazepine combined\vadjust{\goodbreak}
with venlafaxine (determined by preference). Thus, although the study
involved randomization, it is observational with respect to the
treatment options switch or augment. For a complete description see
\citet{Rush}; see Section~A.10 of the supplemental article
[\citet{Schulte}] for a schematic of the design.

To demonstrate formulation of this problem within the framework of
Sections~\ref{s:framework} and \ref{s:defining}, we take level 2A to
be part of level 2 and consider only levels 2 and 3, calling them
stages (decision points) 1 and 2, respectively ($K=2$). Some patients
in stage 1 without adequate response dropped out of the study without
continuing to stage 2. Hence, we analyze complete case data, excluding
dropouts, from 795 patients entering stage 1; 330 of these
subsequently continued to stage~2.
Let $A_k$, $k=1,2$, be the treatment at stage $k$, taking values 0
(augment) or 1 (switch); both options are feasible for all eligible
subjects. Let $S_{10}$ denote baseline (study entry) QIDS score and
$S_{11}$ denote the most recent QIDS score at the beginning of stage
1, respectively, so that $S_1=(S_{10},S_{11})^T$ is information
available immediately prior to the first decision. Similarly, let
$S_2$ be the information available immediately prior to stage 2; here,
$S_2$ is the most recent QIDS score at the end of stage 1/beginning of
stage 2. Finally, let $T$ be the QIDS score at the end of stage 2.
Because some patients exhibited adequate response at the end of stage
1 and did not progress to stage 2, we define the outcome of interest
to be $-S_2$ (negative QIDS score at the end of stage 1) for patients
not moving to stage 2 and $-(S_2+T)/2$ (average of negative QIDS
scores at the end of stages 1 and 2) otherwise. Thus, writing $L_0 =
\max(5,S_{10}/2)$, $Y = -S_2 I(S_2\leq L_0) -(S_2+T)I(S_2>L_0)/2$, the
cumulative average negative QIDS score. Thus, this demonstrates the
case where outcome is a function of accrued information over the
sequence of decisions.\looseness=-1

%
\begin{table*}[t]
\caption{STAR*D data analysis results. Asterisks indicate
evidence at level of significance 0.05 (0.10) that the main effect
(treatment contrast) parameter is nonzero}
\label{t:stard2}
\begin{tabular*}{\textwidth}{@{\extracolsep{\fill}}ld{2.2}ccd{2.2}cc@{}}
\hline
& \multicolumn{3}{c}{\textbf{Q-learning}} &  \multicolumn{3}{c}{\textbf{A-learning}}\\[-4pt]
& \multicolumn{3}{c}{\hrulefill}  & \multicolumn{3}{c@{}}{\hrulefill}\\
\textbf{Parameter} & \multicolumn{1}{c}{\textbf{Estimate}} & $\bolds{95\%}$ \textbf{CI} &
$\bolds{p}$\textbf{-value} & \multicolumn{1}{c}{\textbf{Estimate}} & $\bolds{95\%}$ \textbf{CI} &
$\bolds{p}$\textbf{-value}\\
\hline
\multicolumn{7}{c}{Stage 2}                                                                                         \\
$\beta_{20}$                & -1.46 & ($-3.47, 0.55$)  &   & -1.47           & ($-3.49, 0.54$)  &   \\
$\beta_{21}$                & -0.75 & ($-0.88, -0.61$) & * & -0.75           & ($-0.88, -0.61$) & * \\
$\beta_{22}$                & 1.17  & (0.52, 1.81)      & * & 1.17 & (0.52, 1.81)      & * \\
$\psi_{20}$                 & 1.10  & (0.02, 2.19)      & * & 1.12 & (0.03, 2.22)      & * \\[6pt]
\multicolumn{7}{c}{Stage 1}                                                                             \\
$\beta_{10}$                & -0.62 & ($-1.94, 0.70$)  &   & -0.30           & ($-1.69, 1.0$9)  &   \\
$\beta_{11}$                & -0.54 & ($-0.62, -0.45$) & * & -0.55           & ($-0.64, -0.4$6) & * \\
$\beta_{12}$                & -0.08 & ($-0.60, 0.45$)  &   & 0.10 & ($-0.46, 0.66$)  &   \\
$\psi_{10}$                 & 1.11  & (0.28, 1.94)      & * & 0.73 & ($-0.18, 1.65$)  &   \\
$\psi_{11}$                 & 1.02  & ($-0.08 , 2.11$)  & * & 0.44 & ($-0.83, 1.72$)  &       \\
\hline
\end{tabular*}
\end{table*}

From (\ref{eq:qK}),
$Q_2(\bar{s}_2,\bar{a}_2) = E(Y|\barS_2=\bar{s}_2,\barA_2=\bar
{a}_2) =
-s_2\{I(s_2\leq l_0) + I(s_2>l_0)/2\} + E(-T
|\barS_2=\bar{s}_2,\break \barA_2=\bar{a}_2,S_2>l_0) I(s_2>l_0)/2$, so that
$V_2(\bar{s}_2,a_1) =   -s_2 I(s_2\leq l_0) +
\{-s_2+U_2(\bar{s}_2,a_1)\}I(s_2>l_0)/2,$ where $U_2(\bar{s}_2,a_1) =
\max_{a_2} E(-T |\barS_2=\bar{s}_2, A_1={a}_1,A_2=a_2,\break S_2>l_0)$.
Thus, from
(\ref{eq:qk}),
\begin{eqnarray*}
Q_1(s_1,a_1) &= &E \bigl[ -S_2
I(S_2\leq l_0) \\
&&\hspace*{11pt}{}+ \bigl\{-S_2+U_2(
\bar{s}_2,a_1) \bigr\}I(S_2>l_0)/2|\\
&&\hspace*{96pt}S_1=s_1,A_1=a_1 \bigr].
\end{eqnarray*}

We describe implementation for \iQ-learning. At the second decision
point, we must posit a model for $Q_2(\bar{s}_2,\bar{a}_2)$. From the
form of $Q_2(\bar{s}_2,\bar{a}_2)$, we need only\vadjust{\goodbreak} specify a model for $E(-T
|\barS_2=\bar{s}_2,\barA_2=\bar{a}_2,S_2>l_0)$; given the form of the
conditioning set, this may be carried out using only the data from
patients moving to stage 2. Based on exploratory analysis, defining
$s_{22}$ to be the slope of the QIDS score over stage 1 based on
$s_{11}$ and
$s_2$, we
took this model to be of the form $\beta_{20} + \beta_{21}s_2 +
\beta_{22}s_{22} + \psi_{20}a_2$, so that the posited \iQ-function is
%
\begin{eqnarray}\label{eq:q2example}
&&Q_2(\bar{s}_2,\bar{a}_2; \xi_2)
\nonumber\\
&&\quad= -s_2 \bigl\{I(s_2\leq l_0) +
I(s_2>l_0)/2 \bigr\}
\nonumber
\\[-8pt]
\\[-8pt]
\nonumber
&&\qquad{}+ I(s_2>
l_0) \\
&&\qquad{}\times (\beta_{20} + \beta_{21}s_2 +
\beta_{22}s_{22} + \psi_{20}a_2)/2,\nonumber
\end{eqnarray}
$\xi_2=(\beta_{20},\beta_{21},\beta_{22},\psi_{20})^T$. Under
(\ref{eq:q2example}), $V_2(\bar{s}_2,a_1;  \xi_2) = -s_2\{I(s_2\leq
l_0) +
I(s_2>l_0)/2\} + I(s_2> l_0) \times  \{\beta_{20} + \beta_{21}s_2 +
\beta_{22}s_{22} + \psi_{20} I(\psi_{20}>0)\}/2$, and the
``responses'' $\tilV_{2,i}$ for use in (\ref{eq:qlearnk}) may then be
formed by substituting the estimate for $\xi_2$. Based on exploratory
analysis, we took the posited \iQ-function at the first stage to be
$Q_1(s_1,a_1;\xi_1) = \beta_{10} + \beta_{11}s_{11} + \beta_{12}s_{12}
+ a_1( \psi_{10} + \psi_{11}s_{12})$, where $s_{12}$ is the slope of the
QIDS score prior to stage 1 based on $s_{10}$ and $s_{11}$, and
$\xi_1=(\beta_{10},\beta_{11},\beta_{12},\psi_{10},\psi_{11})^T$. For
\iA-learning, we posited models for the functions
$h_k(\bar{s}_k,\bar{a}_{k-1})$ and $C_k(\bar{s}_k,\bar{a}_{k-1})$,
$k=1,2$, in
the obvious way analogous to those above, and we took the
propensity models to be of the form $\pi_{2}(\bar{s}_2, a_1;\phi_2) =
\mathrm{expit}(\phi_{20} + \phi_{21}s_2 + \phi_{22}s_{22} +
\phi_{23}a_1)$ and $\pi_{1}(s_1;\phi_1) = \mathrm{expit}(\phi_{10} +
\phi_{11}s_{11} + \phi_{12}s_{12})$. Section~A.11 of the
supplemental article [\citet{Schulte}] presents model diagnostics.

The results are given in Table~\ref{t:stard2}. To describe
implementation, we consider interactions significant based on a test
at level $\alpha=0.10$. At the first stage, \iQ-learning suggests a
treatment switch for those with QIDS slope prior to stage 1 greater
than $-1.09$ (obtained by solving $1.11+1.02 S_{12} = 0$); \iA-learning
assigns a treatment switch for those with this QIDS slope greater than
$-1.66$.
At stage 2, the results suggest that all patients
should switch and not augment their existing treatments.

\section{Discussion}
\label{s:discuss}

We have provided a self-contained account of \iQ- and \iA-learning
methods for estimating optimal dynamic treatment regimes, including a
detailed discussion of the underlying statistical framework in which
these methods may be formalized and of their relative merits. Our
discussion of \iA-learning is limited to the case of two treatment
options at each decision. Our simulation studies suggest that, while
\iA-learning may be inefficient relative to \iQ-learning in estimating
parameters that define the optimal regime when the \iQ-functions
required for the latter are correctly specified, \iA-learning may
offer robustness to such misspecification. Nonetheless, \iQ-learning
may have practical advantages in that it involves modeling tasks
familiar to most data analysts, allowing the use of standard
diagnostic tools. On the other hand, \iA-learning may be preferred in
settings where it is expected that the form of the decision rules
defining the optimal regime is not overly complex. However,
\iA-learning increases in\vadjust{\goodbreak} complexity with more than two treatment
options at each stage, which may limit its appeal. Interestingly,
in the simulation scenarios we consider, inefficiency and bias in
estimation of
parameters defining the optimal regime does not necessarily translate
into large degradation of average performance of the estimated regime
for either method.

Although our simple simulation studies provide some insight into the
relative merits of these methods, there remain many unresolved issues
in estimation of optimal treatment regimes. Approaches to address the
challenges of high-dimensional information and large numbers of
decision points are required. Existing methods for model selection
focusing on minimization of prediction error may not be best for
developing models optimal for decision-making. When $K$ is very
large, the number of parameters in the models required for \iQ- and
\iA-learning becomes unwieldy. The analyst may wish to postulate
models in which parameters are shared across decision points; see
\cite{Robins04}, \citet{Orellana08}, \citet{Orellana10} and
\citet{CM}.

In our development, we have invoked a strong version of the sequential
randomization assumption to simplify supporting arguments.
\citet{RichardsonRobins} allow identification of potential outcomes
under possibly weaker assumptions via graphical representations.
These authors also extend the notion of a dynamic treatment regime.

Formal inference procedures for evaluating the uncertainty associated
with estimation of the optimal regime are challenging due to the
nonsmooth nature of decision rules, which in turn leads to
nonregularity of the parameter estimators; see \citet{Robins04},
\citet{Chakraborty}, \citet{Laberetal}, \citet{MoodieRichardson},
\citet{Song} and \citet{LaberMurphy}.

We have discussed sequential decision-making in the context of
personalized medicine, but many other applications exist where, at one
or more times in an evolving process, an action must be taken from
among a set of plausible actions. Indeed, \iQ-learning was originally
proposed in the computer science literature with these more general
problems in mind; see \citet{Shortreed}.


\section*{Acknowledgments}
Supported in part by NIH Grants R37 AI031789, R01 CA051962, R01
CA085848, P01 CA142538 and T32 HL079896.


\begin{supplement}[id=suppA]
\stitle{Supplement to ``Q- and A-Learning Methods for Estimating
Optimal Dynamic Treatment Regimes''}
\slink[doi]{10.1214/13-STS450SUPP} 
\sdatatype{.pdf}
\sfilename{sts450\_supp.pdf}
\sdescription{Due to space constraints, technical details and further
results are given in the supplementary document \citet{Schulte}.}
\end{supplement}


%

\end{document}